\documentclass[12pt]{article}

\usepackage{multirow} 
\usepackage{fullpage}
\usepackage{latexsym}              
\usepackage{amssymb}               
\usepackage{amsmath}               
\usepackage{amsthm}                
\usepackage{bbm}
\usepackage{url}
\usepackage{rotating}
\usepackage{caption}
\usepackage{setspace}
\usepackage{thmtools}
\usepackage{color}
\usepackage{algpseudocode}
\usepackage[ruled]{algorithm}
\usepackage{algorithmicx}

\usepackage{arydshln}

\usepackage{subcaption}
\usepackage{graphicx}
\usepackage[english]{babel}

\usepackage{lscape}  
\usepackage{algorithm}
\usepackage{algpseudocode}
\usepackage[toc,page]{appendix}

\newcommand{\blind}{1}

\declaretheoremstyle[]{normalhead}

\newcommand{\by}{{\bf y}}

\newcommand{\bt}{{\bf t}}
\newcommand{\bz}{{\bf z}}

\newcommand{\bI}{{\bf I}}

\newcommand{\bX}{{\bf X}}

\newcommand{\bzero}{{\bf 0}}

\newcommand{\bbeta}{{\boldsymbol\beta}}

\newcommand{\beq}{\begin{equation}}
\newcommand{\eeq}{\end{equation}}

\makeatletter

\newcount\mn \newcount\hr
\def\timeofday{%
  \hr=\time \divide\hr 60
  \mn=-\hr \multiply\mn 60 \advance\mn \time
  \ifnum\hr=0%
     {12\,:\,\twodigits\mn\,am}%
  \else{%
     \ifnum\hr<12%
        {\number\hr\,:\,\twodigits\mn\,am}%
     \else{%
        \ifnum\hr=12%
          {\number\hr\,:\,\twodigits\mn\,pm}%
        \else%
          {\advance\hr -12 \number\hr\,:\,\twodigits\mn\,pm}%
        \fi}%
     \fi}%
  \fi}
\def\twodigits#1{\ifnum #1<10 0\fi \number#1}
\def\tersetoday{\ifcase\month\or
  Jan.\or Feb.\or Mar.\or Apr.\or May\or June\or
  July\or Aug.\or Sep.\or Oct.\or Nov.\or Dec.\fi
  \space\number\day, \number\year}

\date{\today}

\begin{document}

\title{\bf Efficient and Robust Block Designs for Order-of-Addition Experiments}

\if1\blind
{
   \author{Chang-Yun Lin\footnote{\color{black}Postal address: 250 Kuo-Kuang Rd., Taichung 402, Taiwan; Tel: +886-4-22860133 ext. 607; E-mail: chlin6@nchu.edu.tw}\\   \small \it Department of Applied Mathematics and Institute of Statistics,\\ \small \it National Chung Hsing University, Taichung, Taiwan, 40227\\  
 }\fi
    
\date{} 
\maketitle

\begin{abstract}

Designs for Order-of-Addition (OofA) experiments have received growing attention due to their impact on responses based on the sequence of component addition. In certain cases, these experiments involve heterogeneous groups of units, which necessitates the use of blocking to manage variation effects. Despite this, the exploration of block OofA designs remains limited in the literature. As experiments become increasingly complex, addressing this gap is essential to ensure that the designs accurately reflect the effects of the addition sequence and effectively handle the associated variability. 
Motivated by this, {\color{black}the study} seeks to address the gap by expanding the indicator function framework for block OofA designs. 
{\color{black}The word length pattern is proposed} as a criterion for selecting robust block OofA designs. To improve search efficiency and reduce computational demands, {\color{black}an algorithm is developed} that employ orthogonal Latin squares for design construction and selection, thereby minimizing the need for exhaustive searches.
{\color{black}The} analysis, supported by correlation plots, reveals that the algorithms effectively manage confounding and aliasing between effects. Additionally, simulation studies indicate that designs based on {\color{black}the} proposed criterion and algorithms achieve power and type I error rates comparable to those of full block OofA designs. This approach offers a practical and efficient method for constructing block OofA designs and may provide valuable insights for future research and applications.

\end{abstract}
 

\section{Introduction}\label{se:intro}
{\color{black}Designs for Order-of-Addition (OofA) experiments have received increasing attention in recent years.} 
In these designs, the response is influenced by the sequence in which components or materials are added. 
An early example of an OofA experiment is Fisher’s (1937) study, where a lady claimed she could discern by taste whether the milk or the tea was added first to the cup. 
Such experiments have since been applied in many areas, including medical research (Black et al., {\color{black}2001}), biochemistry (Preuss et al., 2009), measurement science (PerkinElmer, 2016), and task arrangements (Voutchkov et al., 2005; Wilson et al., 2018).
{\color{black}In pharmaceutical manufacturing, for instance, the sequence and timing of adding active ingredients and stabilizers can influence both the efficacy and shelf life of a drug. In materials science, the order of introducing monomers and initiators can alter key polymer characteristics such as strength, flexibility, and durability. Consequently, understanding and optimizing the order of addition is crucial for improving product quality and process efficiency across a wide range of scientific and industrial applications.}

Several models have been proposed in the literature for analyzing data from OofA experiments. 
The most widely used model is the pairwise ordering (PWO) model, first proposed by Van Nostrand (1995), in which the response is fully accounted for by the pairwise ordering of the components. 
Design constructions under the PWO model have been extensively discussed and developed, often utilizing the D criterion to search for optimal OofA designs. 
Notable contributions include the works of Voelkel (2019), Voelkel and Gallagher (2019), Lin and Peng (2019), Mee (2020), Chen et al. (2021), Tsai (2022), Wang and Mee (2022), {\color{black}Zhao et al. (2022), and Xiao et al. (2024).} 
These studies have significantly advanced the methodology for crafting efficient OofA designs under the PWO model, improving the capacity to analyze the influence of component order on response variables.

In contrast to the PWO model, which considers the relative positions of each pair of components, Yang et al. {\color{black}(2021)} introduced the component-position (CP) model, which emphasizes the absolute positions of the components. 
They introduced the concept of component orthogonal arrays (COA), which exhibit high D-efficiency for estimating parameters of the CP model. 
Stokes and Xu (2022) further advanced OofA designs by presenting them with position matrices instead of component matrices. 
Building on these position matrices, they introduced the position-based (PB) approach and developed position models.
This approach significantly reduces the number of parameters required in position models compared to the PWO and CP models.

{\color{black}In many situations, OofA experiments may not be conducted with a homogeneous group of units. These experiments might involve diverse patient groups or be performed in different environments. As a result, variations arising from group-to-group or environment-to-environment differences can influence the analysis outcomes of the OofA experiment.
For instance, in biochemistry, environmental factors such as humidity or equipment variability can influence experimental results. In measurement science, differences in calibration, maintenance, or performance between different instruments can introduce variations in measurement data. 
To mitigate the influence of these variations, blocking is often employed. This involves organizing units into blocks such that the variation within each block is notably smaller than the variation observed between blocks. By doing so, the effects of the blocks are controlled, allowing for a clearer comparison of the effects of the components.
A well-recognized real-world application of block OofA experiments is the five-drug study in the pharmaceutical industry described in Mee (2020). This study utilized two blocks of 20-run drug sequences to mitigate batch-to-batch variations. Wang et al. (2020) also conducted a similar experiment using three batch blocks of 20 runs each. These studies highlight the essential role of block OofA designs in real-world applications. However, methods for constructing such block designs and the criteria for their evaluation remain largely unexplored in the literature.}

As experiments become increasingly complex, addressing this gap is essential to ensure that designs accurately {\color{black}capture} the effects of the addition sequence and effectively handle associated variability. Motivated by this, this paper {\color{black}aims} to {\color{black}bridge} the gap by {\color{black}leveraging} position models to introduce a suitable criterion for assessing block OofA designs and developing a method for constructing highly efficient and robust block OofA designs based on the proposed criterion.
{\color{black}To achieve this, we utilize the indicator function introduced by Cheng and Ye (2004), originally developed for factorial designs. We find that position models align with the framework of the indicator function and thus adapt this concept to tailor the indicator function specifically for OofA designs. Additionally, we incorporate block factors into the indicator function and develop word length patterns to guide the selection of optimal block OofA designs.}

The remainder of this article is organized as follows. 
In Section~\ref{se:O}, we create an indicator function based on the framework of position models for OofA designs. 
In Section~\ref{se:bO}, we extend the concept from Section~\ref{se:O} to develop an indicator function for block OofA designs and propose a criterion based on this indicator function to evaluate these designs. 
In Section~\ref{se:al}, we present algorithms for selecting optimal block OofA designs using the criterion proposed in Section~\ref{se:bO}. 
Several examples are provided to illustrate the process of obtaining these optimal designs.
{\color{black}In Section~\ref{se:5dru}, we conduct a case study to demonstrate the analysis of the block OofA experiment using the proposed methods and the constructed designs.}
In Section~\ref{se:sim}, we evaluate the performance of the proposed block OofA designs using correlation plots and simulation studies. Section~\ref{se:con} provides the concluding remarks.

\section{Position models and indicator function for OofA designs}\label{se:O}

We employ the position models proposed by Stokes and Xu (2022) for OofA designs, as these models involve fewer parameters compared to the PWO and CP models. 
In this section, we first review OofA designs and position models, and then establish a framework for the indicator function for OofA designs, adapted from the framework proposed by Cheng and Ye (2004) for factorial designs.
The modification of the indicator function for block OofA designs will be developed in the Section~\ref{se:bO}.
{\color{black}A notation glossary for the OofA and the proposed block OofA design framework is provided in Table~\ref{tb:notation} in the Appendix.}

\subsection{The OofA designs}\label{se:pb}

\begin{table}							
\centering							
\caption{Two 3-component OofA designs with 6 runs}\label{tb:d3}							
\begingroup							
\begin{subtable}[t]{.3\linewidth}							
\centering							
\caption{$D_1$}\label{tb:d31}							
\begin{tabular}{cccc}							
\hline							
Run	&	$Z_1$	&	$Z_2$	&	$Z_3$	\\
\hline							
1	&	1	&	2	&	3	\\
2	&	1	&	3	&	2	\\
3	&	2	&	1	&	3	\\
4	&	2	&	3	&	1	\\
5	&	3	&	1	&	2	\\
6	&	3	&	2	&	1	\\
\hline							
\end{tabular}							
\end{subtable}							
\begin{subtable}[t]{.3\linewidth}							
\centering							
\caption{$D_2$}\label{tb:d32}							
\begin{tabular}{cccc}							
\hline							
Run	&	$Z_1$	&	$Z_2$	&	$Z_3$	\\
\hline							
1	&	1	&	2	&	3	\\
2	&	1	&	2	&	3	\\
3	&	2	&	1	&	3	\\
4	&	3	&	1	&	2	\\
5	&	3	&	1	&	2	\\
6	&	3	&	2	&	1	\\
\hline							
\end{tabular}							
\end{subtable}							
\endgroup							
\end{table}

Building upon the methodology outlined by Stokes and Xu (2022), we define $m$-component OofA designs as follows. 
Let the factor $Z_j$ represent the experimental position of the $j$th component in the OofA experiment, where $j=1, 2, \cdots, m$. 
We refer to these factors as position factors.
Each factor comprises $m$ levels, $\{1, 2, \cdots, m\}$, and within a single run (unit), the level combination {\color{black}$\bz=(z_1, z_2, \cdots, z_m)$} of the $m$ factors forms a permutation of the positions $\{1, 2, \cdots, m\}$.
{\color{black}Note that $Z_j$ (uppercase) refers to the $j$th factor, while $z_j$ (lowercase) refers to the $j$th element of $\bz$.}
Table~\ref{tb:d3} presents two 3-factor OofA designs with 6 runs. Design $D_1$ is a full OofA design, while $D_2$ is a fractional OofA design that includes two replicated design points: $(1, 2, 3)$ and $(3, 1, 2)$.
{\color{black}For the design point $\bz = (3, 1, 2)$, components 1, 2, and 3 are assigned to experimental positions 3, 1, and 2, respectively.
Thus, the corresponding sequence is component~2 $\rightarrow$ component~3 $\rightarrow$ component~1.}
Note that in the OofA designs discussed in this paper, the $j$th factor indicates the position of the $j$th component. This differs from the traditional representation, where the $j$th factor indicates the component specified for the $j$th position.

Since positions inherently possess an order, Stokes and Xu (2022) investigated their effects using polynomial functions. 
They defined the orthogonal polynomial contrasts of degrees 1 and 2 over the set of positions as $p_1(z) = h_1 \left( z - \frac{m+1}{2} \right)$ and $p_2(z) = h_2 \left[ \left( z - \frac{m+1}{2} \right)^2 - \frac{m^2-1}{12} \right]$, respectively.
Here, $h_1$ and $h_2$ are scalars ensuring that the length of each contrast vector is $\sqrt{m}$.
Based on this polynomial framework, the following full models can be considered:
\begin{itemize}
\item[(1)] Full first-order position model: $E(y_i)=\beta_0+\sum_{j=1}^{m}\beta_jp_1(z_{ij})$,
\item[(2)] Full quadratic position model: $E(y_i)=\beta_0+\sum_{j=1}^{m}\beta_jp_1(z_{ij})+\sum_{j=1}^{m}\beta_{jj}p_2(z_{ij})$,
\item[(3)] Full second-order position model: $E(y_i)=\beta_0+\sum_{j=1}^{m}\beta_jp_1(z_{ij})+\sum_{j=1}^{m}\beta_{jj}p_2(z_{ij})+\sum_{1\leq j<l\leq m}\beta_{jl}p_1(z_{ij})p_1(z_{il})$.
\end{itemize}
Since the constraints $\sum_{l=1}^m p_l(z) = 0$ and $\sum_{l=1}^m p_l^2(z) = m$, where $l=1,2$, render effects in the full model non-estimable, Stokes and Xu (2022) addressed this issue by omitting several terms from the full models.
This led to the following corresponding reduced models: the reduced first-order position model $E(y_i)=\beta_0+\sum_{j=1}^{m-1}\beta_jp_1(z_{ij})$, the reduced quadratic position model $E(y_i)=\beta_0+\sum_{j=1}^{m-1}\beta_jp_1(z_{ij})+\sum_{j=1}^{m-1}\beta_{jj}p_2(z_{ij})$, and the reduced second-order position model $E(y_i)=\beta_0+\sum_{j=1}^{m-1}\beta_jp_1(z_{ij})+\sum_{j=1}^{m-2}\beta_{jj}p_2(z_{ij})+\sum_{1\leq j<l\leq m-1}\beta_{jl}p_1(z_{ij})p_1(z_{il})$.

\subsection{Indicator function for OofA designs}\label{se:id}

Cheng and Ye (2004) introduced the indicator function primarily for factorial designs. 
{\color{black}Unlike a conventional indicator function that takes only 0 or 1 values, the indicator function defined by the authors outputs the number of times a level combination appears in a design. For example, if a level combination does not appear in a design, the indicator function outputs 0; if it appears twice, the indicator function outputs 2, and so on.}
Since position models align with the framework of the indicator function, we adapt this concept to modify the indicator function specifically for OofA designs based on position models.

Let $\mathcal{O}$ denote a full $m$-component OofA design, encompassing all possible $m!$ design points. 
{\color{black}For each factor, we} define a set of orthogonal polynomial contrasts $p_0(z), p_1(z), \cdots, p_{m-1}(z)$ {\color{black}for any scalar $z\in\{1,\cdots,m\}$}, satisfying:
\begin{equation}\label{eq:pu}
\sum_{z\in\{1,\cdots,m\}}p_u(z)p_v(z)=\left\{
\begin{array}{cl}
0&\mbox{if }u\neq v,\\
m&\mbox{if }u=v.
\end{array}
\right.
\end{equation}

As an illustration, orthogonal polynomial contrasts of degrees 0 to 4 are derived as follows: mean contrast: $p_0(z) = 1$, linear contrast: $p_1(z) = h_1 \left(z - \frac{m+1}{2}\right)$, quadratic contrast: $p_2(z) = h_2 \left[ \left(z - \frac{m+1}{2}\right)^2 - \frac{m^2 - 1}{12} \right]$, cubic contrast: $p_3(z) = h_3 \left[ \left(z - \frac{m+1}{2}\right)^3 - \left(z - \frac{m+1}{2}\right) \left(\frac{3m^2 - 7}{20}\right) \right]$, and quartic contrast: $p_4(z) = h_4 \left[ \left(z - \frac{m+1}{2}\right)^4 - \left(z - \frac{m+1}{2}\right)^2 \left(\frac{3m^2 - 13}{14}\right) + \frac{3(m^2 - 1)(m^2 - 9)}{560} \right]$.
The definitions of $p_1(z)$ and $p_2(z)$ align with those provided by Stokes and Xu (2022).
For instance, with $ m=3 $, the orthogonal polynomial contrasts are $ (p_0(1),p_0(2),p_0(3))=(1,1,1) $, $ (p_1(1),p_1(2),p_1(3))=\sqrt{\frac{3}{2}}(-1,0,1) $, and $ (p_2(1),p_2(2),p_2(3))=\sqrt{\frac{1}{2}}(1,-2,1) $.

Let $T = \{0, 1, \cdots, m-1\}$ denote a set of polynomial degrees, and let ${\cal T} = T^m$ represent the Cartesian product of $m$ copies of $T$.
We define the polynomial term for a design point $\bz = (z_1, z_2, \cdots, z_m)$ corresponding to $\bt = (t_1,t_2, \cdots, t_m) \in {\cal T}$ as follows:
\begin{equation} 
X_{\bt}(\bz) = \left\{
\begin{array}{ll}
\prod_{j=1}^m p_{t_j}(z_j),&\mbox{if }\bz\in \cal O,\\
0,&\mbox{otherwise}.
\end{array}
\right.
\end{equation} 
{\color{black}For $\bt = (t_1, t_2, \cdots, t_m)$, $t_j = 0$ indicates that the effect of $Z_j$ is not involved, $t_j = 1$ implies that the linear effect of $Z_j$ is involved, $t_j = 2$ implies that the quadratic effect of $Z_j$ is involved, and so on.
A $\bt$ contains information about factor effects.
For instance, when $m=3$, $\bt = 010$ (simplified from $\bt = (0,1,0)$) indicates the linear effect of $Z_2$, while $\bt = 212$ indicates the interaction of the quadratic effect of $Z_1$, the linear effect of $Z_2$, and the quadratic effect of $Z_3$.}
The polynomial term for the design point $\bz = 312$ (simplified from $\bz = (3,1,2)$), corresponding to $\bt = 212$, is computed as $X_{212}(312) = p_2(3) \times p_1(1) \times p_2(2) = \left(\sqrt{\frac{1}{2}}\right) \times \left(-\sqrt{\frac{3}{2}}\right) \times \left(-2\sqrt{\frac{1}{2}}\right) = \sqrt{\frac{3}{2}}$.
In particular, for any $ \bz \in {\cal O} $, $ X_\bzero(\bz)=\prod_{j=1}^m p_0(z_j)=\prod_{j=1}^m 1=1 $, where $ \bzero=(0,0,\cdots,0) $. Thus, $ X_\bzero(\bz) $ is referred to as the constant term.

Let $ \cal D $ be an $ m $-component OofA design within the design space $ \cal O $, meaning that for every $ \bz \in \cal D $, $ \bz \in {\cal O} $.
The indicator function for design $ \cal D $ is defined as follows:
\begin{equation} 
F_{\cal D}(\bz) = \sum_{\bt \in {\cal T}} a_{\bt} X_\bt(\bz),
\end{equation} 
where the coefficient of the term $ X_\bt(\bz) $ is computed as:
\begin{equation} 
a_\bt = \frac{1}{m^m} \sum_{\bz \in \cal D} X_\bt(\bz).
\end{equation} 
In particular, $ a_\bzero = \frac{n}{m^m} $, where $ n $ is the run size of $ \cal D $. The indicator function $ F_{\cal D}(\bz) $ is undefined when $ \bz \notin {\cal O} $.

{\color{black}In the indicator function, $a_\bt$ measures the severity of aliasing between the mean and the effects in $\bt$. 
Generally, a smaller $|a_\bt|$ implies a lower degree of aliasing. 
When $a_\bt = 0$, the mean and the effects in $\bt$ are not aliased. 
The severity of aliasing can be standardized using $\left(\frac{a_\bt}{a_\bzero}\right)^2$, which ranges from 0 (indicating no aliasing) to 1 (indicating complete aliasing).}

For example, the indicator functions of the two 3-component OofA designs $ D_1 $ and $ D_2 $ in Table~\ref{tb:d3} are:
\begin{equation} 
\begin{aligned}
F_{D_1}(\bz) = & 0.22 X_{000}(\bz) - 0.11 X_{110}(\bz) - 0.11 X_{220}(\bz) - 0.11 X_{101}(\bz) - 0.11 X_{011}(\bz) \\
& + 0.16 X_{211}(\bz) + 0.16 X_{121}(\bz) - 0.11 X_{202}(\bz) + 0.16 X_{112}(\bz) - 0.11 X_{022}(\bz) - 0.16 X_{222}(\bz),
\end{aligned}
\end{equation} 
and
\begin{equation} 
\begin{aligned}
F_{D_2}(\bz) = & 0.22 X_{000}(\bz) + 0.05 X_{100}(\bz) + 0.08 X_{200}(\bz) - 0.14 X_{010}(\bz) - 0.11 X_{110}(\bz) - 0.08 X_{020}(\bz) \\
& + 0.13 X_{120}(\bz) - 0.11 X_{220}(\bz) + 0.09 X_{001}(\bz) - 0.17 X_{101}(\bz) - 0.03 X_{201}(\bz) - 0.06 X_{011}(\bz) \\
& + 0.08 X_{211}(\bz) - 0.03 X_{021}(\bz) + 0.24 X_{121}(\bz) - 0.09 X_{221}(\bz) - 0.16 X_{102}(\bz) - 0.06 X_{202}(\bz) \\
& + 0.1 X_{012}(\bz) + 0.16 X_{112}(\bz) + 0.14 X_{212}(\bz) - 0.17 X_{022}(\bz) - 0.05 X_{122}(\bz) - 0.16 X_{222}(\bz).
\end{aligned}
\end{equation} 
One can verify that for $ \bz = 123 $, $ F_{D_1}(123) = 1 $ and $ F_{D_2}(123) = 2 $, while for $ \bz = 132 $, $ F_{D_1}(132) = 1 $ and $ F_{D_2}(132) = 0 $. 
{\color{black}Note that $a_{100} = 0$ in $F_{D_1}(\bz)$; therefore, the linear effect of $Z_1$ in $D_1$ is not aliased with the mean.}

\subsection{Word length pattern for OofA designs}\label{se:wd}

In the indicator function, $\bt$ is termed a ``word" if $a_\bt \neq 0$. 
Let $||\bt|| = \sum_{j=1}^m t_j$, which {\color{black}is} the polynomial degree of $\bt$. 
For example, $||\bt|| = 1$ when $\bt = 010$, and $||\bt|| = 4$ when $\bt = 121$.
According to the effect hierarchy principle (see Wu {\color{black}and} Hamada, {\color{black}2021}), effects with lower polynomial degrees are considered more important than those with higher polynomial degrees, and effects with the same polynomial degree are regarded as equally important. Therefore, for two words $\bt_1$ and $\bt_2$, $\bt_1$ is considered more important than $\bt_2$ if $||\bt_1|| < ||\bt_2||$, and they are considered equally important if $||\bt_1|| = ||\bt_2||$.

Drawing on the effect hierarchy principle and utilizing $\left(\frac{a_\bt}{a_\bzero}\right)^2$ as a standardized measure for aliasing severity between the mean and the effects in $\bt$, the word length pattern (WLP) can be defined as:
\begin{equation} 
W = (w_{1}, w_{2}, \cdots, w_{m(m-1)}),
\end{equation} 
where $w_{l} = \sum_{||\bt||=l} \left(\frac{a_\bt}{a_\bzero}\right)^2$ for $l=1,2,\cdots,m(m-1)$.
For two OofA designs, ${\cal D}_1$ and ${\cal D}_2$, if $w_l({\cal D}_1) = w_l({\cal D}_2)$ for $l = 1, \cdots, j-1$ and $w_j({\cal D}_1) < w_j({\cal D}_2)$, then ${\cal D}_1$ is considered to have less aberration than ${\cal D}_2$ and is regarded as superior. This relationship between the word length patterns of the two designs is denoted as $W({\cal D}_1) \ll W({\cal D}_2)$.
As addressed by Cheng et al. (2002), Mandal and Mukerjee (2005), and Stokes and Xu ({\color{black}2022}), minimum aberration designs are highly efficient and robust under model uncertainty.

For example, the WLPs for $ D_1 $ and $ D_2 $ in Table~\ref{tb:d3} are
\begin{equation} 
W(D_1) = (0, 0.75, 0, 2.25, 0, 0.5)
\end{equation} 
and
\begin{equation} 
W(D_2) = (0.58, 1.13, 1.08, 2.63, 0.58, 0.5).
\end{equation} 
Comparing the two WLPs, we find that $w_1(D_1) = 0 < w_1(D_2) = 0.58$, indicating that $W(D_1) \ll W(D_2)$. 
This result shows that $D_1$ has less aberration than $D_2$ and is therefore more desirable.

{\color{black}The resolution of a design can be simply defined as the smallest index $i$ such that $w_{i} > 0$. In practice, designs with the highest resolution can be considered first, and the minimum aberration criterion can then be applied to select the most desirable design from among those with the highest resolution.}

\section{Position models and indicator function for block OofA designs}\label{se:bO}
Based on the framework for OofA designs introduced in Section~\ref{se:O}, we extend the models, indicator function, and WLP to block OofA designs in this section.

\subsection{Block OofA designs}\label{se:bpb}

\begin{table}									
\centering									
\caption{Two 3-component block OofA designs with 2 blocks size 3}\label{tb:bd3}									
\begingroup									
\begin{subtable}[t]{.35\linewidth}									
\centering									
\caption{$D_1'$}\label{tb:bd31}									
\begin{tabular}{ccccc}									
\hline									
Run	&	$Z_1$	&	$Z_2$	&	$Z_3$	&	$B$	\\
\hline									
1	&	1	&	2	&	3	&	1	\\
2	&	1	&	3	&	2	&	2	\\
3	&	2	&	1	&	3	&	1	\\
4	&	2	&	3	&	1	&	2	\\
5	&	3	&	1	&	2	&	1	\\
6	&	3	&	2	&	1	&	2	\\
\hline									
\end{tabular}									
\end{subtable}									
\begin{subtable}[t]{.35\linewidth}									
\centering									
\caption{$D_2'$}\label{tb:bd32}									
\begin{tabular}{ccccc}									
\hline									
Run	&	$Z_1$	&	$Z_2$	&	$Z_3$	&	$B$	\\
\hline									
1	&	1	&	2	&	3	&	1	\\
2	&	1	&	3	&	2	&	2	\\
3	&	2	&	1	&	3	&	2	\\
4	&	2	&	3	&	1	&	1	\\
5	&	3	&	1	&	2	&	1	\\
6	&	3	&	2	&	1	&	2	\\
\hline									
\end{tabular}									
\end{subtable}									
\endgroup									
\end{table}															

Consider an $m$-component block OofA design, denoted as ${\cal D}'$, formed by arranging the design points of an $m$-component OofA design into $k$ blocks. 
The block design ${\cal D}'$ includes $m+1$ factors: the first $m$ factors, $Z_1, Z_2, \cdots, Z_m$, are the position factors as defined in Section~\ref{se:pb}, and the last factor, $B$, is the block factor, with levels $\{1, 2, \cdots, k\}$ indicating the blocks.
Table~\ref{tb:bd3} illustrates two block OofA designs, obtained by arranging the OofA designs $D_1$ and $D_2$ from Table~\ref{tb:d3} into 2 blocks of size 3. 
The design point $\bz' = (2,1,3,1)$ indicates that the component position $(2,1,3)$ is allocated to block 1.

In addition to the orthogonal polynomial contrasts defined in Section~\ref{se:id} for the position factors $Z_1, Z_2, \cdots, Z_m$, we define another set of orthogonal polynomial contrasts $c_0(b), c_1(b), \cdots, c_{k-1}(b)$ for the block factor $B$, which satisfy

\begin{equation}\label{eq:cu}
\sum_{b\in\{1,\cdots,k\}}c_u(b)c_v(b)=\left\{
\begin{array}{cl}
0&\mbox{if }u\neq v,\\
k&\mbox{if }u=v.
\end{array}
\right.
\end{equation}
Note that the orthogonal polynomial contrasts defined in (\ref{eq:cu}) for the block factor $B$ have the same form as those defined in (\ref{eq:pu}) for the position factors $Z_1, Z_2, \cdots, Z_m$. However, unlike position factors, which are numerical, the block factor $B$ is nominal. 
Thus, the contrasts in (\ref{eq:cu}) indicate effect comparisons between blocks. 
For example, for a 3-block design ($k=3$), the contrast $(c_1(1), c_1(2), c_1(3)) = \sqrt{\frac{3}{2}}(-1, 0, 1)$ compares the block effect between block 1 and block 3. The contrast $(c_2(1), c_2(2), c_2(3)) = \sqrt{\frac{1}{2}}(1, -2, 1)$ compares the block effect of block 2 with the average of the block effects of blocks 1 and 3. The contrast $(c_0(1), c_0(2), c_0(3)) = (1, 1, 1)$ represents the mean block effect.

For block factorial designs, it is generally assumed that the interactions between the block factor and treatment factors are negligible (see Wu {\color{black}and} Hamada, {\color{black}2021}).
{\color{black}Based on this assumption, the following reduced models for the block OofA designs can be considered, where the estimability constraints $\sum_{l=1}^m p_l(z) = 0$ and $\sum_{l=1}^m p_l^2(z) = m$ are applied: 
\begin{itemize}
\item[($1'$)] Reduced first-order block-position model: $E(y_i)=\beta_0+\sum_{j=1}^{m-1}\beta_jp_1(z_{ij})+\sum_{l=1}^{k-1}\gamma_{l}c_l(b_i)$,
\item[($2'$)] Reduced quadratic block-position model: $E(y_i)=\beta_0+\sum_{j=1}^{m-1}\beta_jp_1(z_{ij})+\sum_{j=1}^{m-1}\beta_{jj}p_2(z_{ij})+\sum_{l=1}^{k-1}\gamma_{l}c_l(b_i)$,
\item[($3'$)] Reduced second-order block-position model: $E(y_i)=\beta_0+\sum_{j=1}^{m-1}\beta_jp_1(z_{ij})+\sum_{j=1}^{m-2}\beta_{jj}p_2(z_{ij})+\sum_{1\leq j<l\leq m-1}\beta_{jl}p_1(z_{ij})p_1(z_{il})+\sum_{l=1}^{k-1}\gamma_{l}c_l(b_i)$.
\end{itemize}
}

{\color{black}When performing data analysis, we employ forward regression analysis with full models instead of reduced models to identify significant effects. By considering full models and not pre-omitting terms, our approach can be more adaptable to real-world experimental settings, where simplifications or assumptions about omitting certain terms might not always hold.}

\subsection{Indicator function for block OofA designs}\label{se:bid}
{\color{black}The indicator function was extended by Lin (2014) for block factorial designs. We modify it to better accommodate the position model in block OofA designs.}
Let $\mathcal{O}'$ denote a full $m$-component block OofA design with $k$ blocks of size $m!$, encompassing all possible $km!$ design points (by assigning the full $m$-component OofA design into each block).
Let $\bz'=(z_1,z_2,\cdots,z_m, b)$ be a level combination of factors $Z_1,Z_2,\cdots,Z_m$, and $B$, representing the arrangement of the component positions $(z_1,z_2,\cdots,z_m)$ in block $b$. 

In addition to the set $T$ defined in Section~\ref{se:id} for the position factors $Z_1,Z_2,\cdots,Z_m$, we define another set $S=\{0,1,\cdots,k-1\}$ for the block factor $B$. 
Let ${\cal T'}=T^m\times S$ and define the polynomial term for $\bz'$ on $\bt'=(t_1,t_2,\cdots,t_m,s)\in{\cal T'}$ as follows:

\begin{equation} 
X_{\bt'}(\bz') = \left\{
\begin{array}{ll}
\prod_{j=1}^m p_{t_j}(z_j)\times c_{s}(b),&\mbox{if }\bz'\in \cal O',\\
0,&\mbox{otherwise},
\end{array}
\right.
\end{equation} 
where $c_s(b)$ for $s=0,1,\cdots,k-1$ are the orthogonal polynomial contrasts defined in Section~\ref{se:bpb} that satisfy (\ref{eq:cu}).  
For example, in the case of $m=3$ and $k=2$, the polynomial term for design point $\bz'=3121$ corresponding to $\bt'=2120$ is computed as $X_{2120}(3121)=p_2(3)\times p_1(1)\times p_2(2)\times c_0(1)=\left(\sqrt{\frac{1}{2}}\right)\times\left(-\sqrt{\frac{3}{2}}\right)\times\left(-2\sqrt{\frac{1}{2}}\right)\times (1)=\sqrt{\frac{3}{2}}$. 
In particular, for any $\bz'\in \cal O'$, $X_{\bzero'}(\bz')=\prod_{j=1}^mp_0(z_j)\times c_0(b)=\prod_{j=1}^m 1\times 1=1$, where $\bzero'=(0,0,\cdots,0,0)$. 
Thus, $X_{\bzero'}(\bz')$ is referred to as the constant term.

Let ${\cal D}'$ be an $m$-component block OofA design in the design space of $\cal O'$, meaning that for every $\bz' \in {\cal D}'$, $\bz' \in \cal O'$.  
The indicator function for the block OofA design ${\cal D}'$ is defined as follows:
\begin{equation} 
F_{{\cal D}'}(\bz')=\sum_{\bt'\in{\cal T'}}a_{\bt'}X_{\bt'}(\bz'),
\end{equation} 
where the coefficient of term $X_{\bt'} (\bz')$ is computed as: 
\begin{equation} 
a_{\bt'}=\frac{1}{km^m}\sum_{\bz'\in {\cal D}'}X_{\bt'}(\bz').
\end{equation} 
In particular, $ a_\bzero' = \frac{n}{km^m} $, where $ n $ is the run size of $ \cal D' $. The indicator function $ F_{\cal D'}(\bz') $ is undefined when $ \bz' \notin {\cal O'} $.

For example, the indicator functions of the two 3-component block OofA designs $D_1'$ and $D_2'$ in Table~\ref{tb:bd3} are 						
\begin{equation} 
\begin{aligned}
F_{D_1'}(\bz')=&0.11X_{0000}(\bz')+0.09X_{0101}(\bz')-0.06X_{1100}(\bz')-0.03X_{2101}(\bz')-0.10X_{1201}(\bz')\\
&-0.06X_{2200}(\bz')-0.09X_{0011}(\bz')-0.06X_{1010}(\bz')+0.03X_{2011}(\bz')-0.06X_{0110}(\bz')\\
&+0.08X_{2110}(\bz')+0.03X_{0211}(\bz')+0.08X_{1210}(\bz')+0.09X_{2211}(\bz')+0.10X_{1021}(\bz')\\
&-0.06X_{2020}(\bz')-0.03X_{0121}(\bz')+0.08X_{1120}(\bz')-0.09X_{2121}(\bz')-0.06X_{0220}(\bz')\\
&-0.08X_{2220}(\bz')\\
\end{aligned}
\end{equation} 
and
\begin{equation} 
\begin{aligned}
F_{D_2'}(\bz')=&0.11X_{0000}(\bz')-0.06X_{1100}(\bz')+0.10X_{2101}(\bz')-0.10X_{1201}(\bz')-0.06X_{2200}(\bz')\\
&-0.06X_{1010}(\bz')-0.10X_{2011}(\bz')-0.06X_{0110}(\bz')+0.08X_{2110}(\bz')+0.10X_{0211}(\bz')\\
&+0.08X_{1210}(\bz')+0.10X_{1021}(\bz')-0.06X_{2020}(\bz')-0.10X_{0121}(\bz')+0.08X_{1120}(\bz')\\
&-0.06X_{0220}(\bz')-0.08X_{2220}(\bz').\\
\end{aligned}
\end{equation} 
One can verify that for $\bz' = 1231$, $F_{D_1'}(1231) = 1$ and $F_{D_2'}(1231) = 1$, while for $\bz' = 2131$, $F_{D_1'}(2131) = 1$ and $F_{D_2'}(2131) = 0$.

\subsection{Word length pattern for block OofA designs}\label{se:bwd}
In the indicator function for block OofA designs, $\bt'= (t_1, t_2, \cdots, t_m, s)$ is referred to as a word if $a_{\bt'}\neq0$. 
The words $ \bt' $ are classified into two types: the pure-type word (denoted as $ w^P $) and the mixed-type word (denoted as $ w^B $). 
A $ \bt' $ is considered a pure-type word if $ s = 0 $, indicating that the block effect is not involved. Conversely, $ \bt' $ is considered a mixed-type word if $ s > 0 $, indicating that the block effect is involved.
A pure-type word represents the aliasing relation between the mean and the position effects in $ \bt' $, as described in Section~\ref{se:id}. 
On the other hand, a mixed-type word demonstrates the confounding relation between the block effect and the position effects in $\bt'$.
For instance, in a 3-component block OofA design with 2 blocks of size 3, the mixed-type word $ \bt' = 1021 $ illustrates the confounding relation between the block effect and the interaction of the linear position effect of $ Z_1 $ and the quadratic position effect of $ Z_3 $.

In the indicator function for block OofA designs, if $ \bt' $ is a pure-type word, the coefficient $ a_{\bt'} $ measures the severity of aliasing between the mean and the interaction of the position effects $ t_1, \cdots, t_m $ in $\bt'$, as described in Section~\ref{se:wd}.
A smaller $|a_\bt'|$ implies a lower degree of aliasing.  
On the other hand, if $\bt'$ is a mixed-type word, coefficient $a_{\bt'}$ measures the severity of confounding between the block effect $s$ and the interaction of the position effects $t_1,\cdots,t_m$ in $\bt'$. 
A smaller $|a_{\bt'}|$ implies a lower degree of confounding. 
When $a_{\bt'}=0$, the block effect $s$ and the interaction of the position effects in $\bt'$ are not confounded. 
In particular, when $s>0$ and $t_1=\cdots=t_m=0$, the coefficient $a_{\bt'}$ measures the severity of confounding between the mean and the block effect $s$.
The severity of aliasing or confounding can be standardized using $\left(\frac{a_{\bt'}}{a_{\bzero'}}\right)^2$, which ranges from 0 (indicating no aliasing or confounding) to 1 (indicating complete aliasing or confounding).

Since the block factor is nominal, all orders of block effects are considered equally important. 
Let $||\bt'||_B=\sum_{j=1}^m t_j$, which calculates the polynomial degree of $(t_1,t_2,\cdots,t_m)$ in $\bt'$. 
For instance, $||\bt'||_B=3$ for $\bt'=1020$ and $\bt'=1021${\color{black}, where $\bt'=1020$ is referred to as a 3rd-order pure-type word and $\bt'=1021$ is referred to as a 3rd-order mixed-type word.}   
For the $l$th-order pure-type words, we define  
\begin{equation} 
w_l^P=\sum_{||{\bt'}||_B=l, s=0}\left(\frac{a_{\bt'}}{a_{\bzero'}}\right)^2,
\end{equation} 
which measures the overall aliasing severity between the mean and the $l$th-order position effects.  
Similarly, for the $l$th-order mixed-type words, we define  
\begin{equation} 
w_l^B=\sum_{||{\bt'}||_B=l, s>0}\left(\frac{a_{\bt'}}{a_{\bzero'}}\right)^2,
\end{equation} 
which measures the overall confounding severity between the block effects and the $l$th-order position effects.
Inspired by Cheng et al. (2004) in the context of block factorial designs, we propose a composite WLP for block OofA designs:  
\begin{equation}\label{eq:W'}
W' = (w_1^P, w_1^B, w_2^P, w_2^B, \cdots, w_{m(m-1)}^P, w_{m(m-1)}^B).
\end{equation}  
The ordering of $w_l^P$ before $w_j^P$ and $w_l^B$ before $w_j^B$ for $l < j$ follows the effect hierarchy principle (Wu and Hamada, 2021) discussed in Section~\ref{se:wd}, which states that lower-order effects are generally more important than higher-order effects.
For the same order $l$, $w_l^P$ is placed before $w_l^B$ because aliasing between position effects and the mean is typically less desirable than confounding between position effects and block effects.
However, this ordering can be adjusted based on specific experimental objectives and the preferences of the experimenters.
The WLP in (\ref{eq:W'}) can be rewritten as 
\begin{equation} 
W'=(w_{(1)}', w_{(2)}',\cdots, w_{(2m(m-1))}'),
\end{equation} 
where $w_{(2(l-1)+1)}'=w_l^P$ and $w_{(2(l-1)+2)}'=w_l^B$ for $l=1,\cdots,m(m-1)$.
For two block OofA designs, $ {\cal D}_1' $ and $ {\cal D}_2' $, if $ w_{(l)}'({\cal D}_1') = w_{(l)}'({\cal D}_2') $ for $ l = 0, \cdots, j-1 $ and $ w_{(j)}'({\cal D}_1') < w_{(j)}'({\cal D}_2') $, then $ {\cal D}_1' $ is considered to exhibit less aberration than $ {\cal D}_2' $ and is regarded as superior.
This relationship between the word length patterns of the two block OofA designs is denoted as $ W'({\cal D}_1') \ll W'({\cal D}_2') $.

For example, the WLPs for the block OofA designs $D_1'$ and $D_2'$ in Table~\ref{tb:bd3} are
\begin{equation} 
W'(D_1')=( 0, 1.33, 0.75, 0, 0, 1.83, 2.25, 0, 0, 1.33, 0.5, 0)
\end{equation} 
and
\begin{equation} 
W'(D_2')=(  0,   0, 0.75,   0,  0, 4.5, 2.25,   0,  0,   0, 0.5,   0).
\end{equation} 
Comparing the two WLPs, we find that $w'_{(1)}(D_1')=w'_{(1)}(D_2')$ and $w'_{(2)}(D'_1) = 1.33 > w'_{(2)}(D'_2)  = 0$, indicating that $ W'(D_2') \ll W'(D_1') $.   
This result shows that $D_2'$ has less aberration than $D_1'$ and is therefore more desirable.

\section{Construction algorithms and examples}\label{se:al}

Since each run of an $m$-component OofA design needs to be a permutation of $\{1, 2, \cdots, m\}$, heuristic algorithms such as the coordinate-exchange (Meyer and Nachtsheim, 1995)  and columnwise-pairwise (Li and Wu, 1997) algorithms used for constructing and searching fractional factorial designs cannot be employed for block OofA designs. 
Although the point-exchange algorithm (Fedorov, 1972; Mitchell, 1974) allows for replacing an entire run with a candidate design point, the extremely large set of candidate points for large $m$ and the need to update the design one run at a time make the search inefficient and challenging for finding good designs.
To reduce computational burden and enhance search efficiency, we propose algorithms that stack candidate 
Latin squares (LSs) to construct block OofA designs. 
{\color{black} Instead of modifying designs by swapping individual runs, our algorithms update designs by exchanging either two LSs or two rows within LSs, significantly improving computational efficiency.} 

Based on the method in 
Stokes and Xu (2022), we first create a set of candidate LSs for $m$-component block OofA designs, where $m > 3$ is a prime or a prime power, as follows. 
First, we generate a Galois field $GF(m)$, denoted as $\{\alpha_0, \alpha_1, \cdots, \alpha_{m-1}\}$, where $\alpha_0$ is the zero element.
Then, we use addition and multiplication defined on $GF(m)$ to construct $m-1$ mutually orthogonal $m \times m$ LSs, denoted as  $L_1, \cdots, L_{m-1}$, with $\alpha_i + \alpha_r \times \alpha_j$ as the $(i,j)$th element in $L_r$ for $r = 1, 2, \cdots, m-1$ and $i, j = 0, 1, \cdots, m-1$.
(Note: Two $m \times m$ LSs are orthogonal if, when superimposed, each pair $(i,j)$ of elements for $i, j = 0, \cdots, m-1$ appears exactly once.)
Since simultaneously permuting columns $3, 4, \cdots, m$ of $L_1, L_2, \cdots, L_{m-1}$ results in another set of $m - 1$ mutually orthogonal $m \times m$ LSs, we perform all $(m-2)!$ possible permutations in lexicographic order for columns $3, 4, \cdots, m$ of $L_1, L_2, \cdots, L_{m-1}$. This procedure generates a set of $(m-1)!$ LSs, which are denoted in a specific order as
\begin{equation}\label{eq:set}
{\color{black}L=\{}L_1, L_2, \cdots, L_{(g-1)(m-1)+f}, \cdots, L_{(m-1)!}{\color{black}\}},
\end{equation}
where $g=1,\cdots,(m-2)!$ and $f=1,\cdots,(m-1)$.
The index $g$ indicates the $g$th permutation performed, and $f$ indicates the $f$th LS in the $g$th group of mutually orthogonal LSs.
These LSs have elements from $0$ to $m-1$. Since the levels of OofA designs range from $1$ to $m$, we add $1$ to all the elements of the LSs to align with the levels of the OofA designs.
For instance, the total 24 LSs for $m=5$ are listed in {\color{black}Appendix Table~\ref{tb:ls}.}

For a given $g$, row-wise concatenating $L_{(g-1)(m-1)+f}$ for $f=1,2,\cdots,m-1$ forms an $m(m-1)\times m$ component orthogonal array (COA, see Yang et al., {\color{black}2021}), denoted as $C_{g}$.
Thus, we can obtain a set of COAs from (\ref{eq:set}), denoted in a specific order as follows:
\begin{equation}\label{eq:setc}
{\color{black}C=\{}C_1,C_2,\cdots,C_g,\cdots,C_{(m-2)!}{\color{black}\}},
\end{equation}
where $C_g=[L_{(g-1)(m-1)+1}^T,L_{(g-1)(m-1)+2}^T,\cdots,L_{(g-1)(m-1)+(m-1)}^T]^T$ for $g=1,2,\cdots,(m-2)!$. 
For instance, for $m=5$, $C_1=[L_1^T,\cdots,L_4^T]^T$, $C_2=[L_5^T,\cdots,L_8^T]^T$, $\cdots$, and $C_6=[L_{21}^T,\cdots,L_{24}^T]^T$, where $L_1,\cdots,L_{24}$ are listed in {\color{black}Appendix Table~\ref{tb:ls}}. 
Further, row-wise concatenating $C_{g}$ for $g=1,\cdots, (m-2)!$ forms a full $m$-component OofA design containing all $m!$ runs.

\begin{algorithm}
\caption{Finding the Optimal Block OofA Design}\label{al:1}
{\color{black}
\begin{algorithmic}[1]
\State Initialize $W'_{\text{opt}} = (999, 999, \dots, 999)$  \Comment{Initial WLP for optimal design}
\State Initialize $d_A', d_B', d_C' \gets \emptyset$ \Comment{Empty designs}

\If{$\lambda > 0$}
    \State Select the first $k\lambda$ COAs from set $C$
    \State Assign $\lambda$ of them to each block to form $d_A'$
    \State Concatenate $d_A'$, $d_B'$, and $d_C'$ to form $D_{opt}'$
\EndIf

\If{$\gamma > 0$ or $\delta > 0$}
    \State Select LSs $L_{k\lambda(m-1)+1},\cdots,L_{k\lambda(m-1)+\left\lceil \frac{k(\gamma m+\delta)}{m} \right\rceil }$ from set $L$
    
    \For{$i = 1$ to $I_1$}
    
        \If{$\gamma > 0$} \Comment{Generate initial design $D_0'$}
            \State Arbitrarily assign $k\gamma$ candidate LSs to $k$ blocks, each containing $\gamma$ LSs, forming $d_B'$
        \EndIf
        \If{$\delta > 0$}
            \State Arbitrarily assign $k\delta$ rows from the remaining LSs to $k$ blocks, each containing $\gamma$ rows, forming $d_C'$
        \EndIf
        \State Concatenate $d_A'$, $d_B'$, and $d_C'$ to form $D_0'$
        \State Calculate WLP of $D_0'$, denoted as $W_0'$
        
        \If{$\gamma > 0$} \Comment{Update $d_B'$}
            \For{$j = 1$ to $I_2$}
                \State Arbitrarily exchange two LSs between blocks in $d_B'$ to form updated $d_B^*$ 
                \State Concatenate $d_A'$, $d_B^*$, and $d_C'$ to form updated design $D_{\text{up}}'$
                \State Calculate WLP of $D_{\text{up}}'$, denoted as $W_{\text{up}}'$
                \If{$W_{\text{up}}' \ll W_0'$}
                    \State $D_0' \gets D_{\text{up}}'$, $W_0' \gets W_{\text{up}}'$, $d_B' \gets d_B^*$
                \EndIf
            \EndFor
        \EndIf 

        \If{$\delta > 0$}
            \For{$k = 1$ to $I_3$} \Comment{Update $d_C'$}
                \State Arbitrarily exchange two rows between blocks in $d_C'$ to form updated $d_C^*$
                \State Concatenate $d_A'$, $d_B'$, and $d_C^*$ to form updated design $D_{\text{up}}'$
                \State Calculate WLP of $D_{\text{up}}'$, denoted as $W_{\text{up}}'$
                \If{$W_{\text{up}}' \ll W_0'$}
                    \State $D_0' \gets D_{\text{up}}'$, $W_0' \gets W_{\text{up}}'$, $d_C' \gets d_C^*$
                \EndIf
            \EndFor
        \EndIf 
          
        \If{$W_0' \ll W_{\text{opt}}'$}
            \State $D_{\text{opt}}' \gets D_0'$, $W_{\text{opt}}' \gets W_0'$
        \EndIf

    \EndFor 
\EndIf

\State \Return $D_{\text{opt}}'$ \Comment{Return the optimal design} 
\end{algorithmic}
}
\end{algorithm}

We utilize the LSs from the set (\ref{eq:set}) and the COAs from the set (\ref{eq:setc}) to construct and search for block OofA designs with less aberration. 
Below are some notations: let $m$ denote the number of components, $k$ denote the number of blocks, and $n_B$ denote the size of each block (number of runs in each block).
Our focus is on designs where all blocks have the same size $n_B$ (balanced block designs, where $w_0^B=0$), and $m$ is a prime or a prime power. 
{\color{black}A pseudocode for the proposed construction algorithm is presented in Algorithm~\ref{al:1}. It can be separately discussed under the following situations.}

\subsection{Block size $n_B$ is not a multiple of $m$}\label{se:r3}
We begin by considering the construction of $m$-component block OofA designs with $k$ blocks, each of size {\color{black}$n_B$ is not a multiple of $m$.}
{\color{black}Let $\left\lfloor x \right\rfloor$ denote the floor function, which yields the largest integer less than or equal to $x$.} When $n_B = 1, 2, \cdots, \left\lfloor \frac{m!}{k} \right\rfloor$ is not a multiple of $m$, it can be expressed as $n_B = \lambda m(m-1) + \gamma m + \delta$, where {\color{black}$\lambda = \left\lfloor \frac{n_B}{m(m-1)} \right\rfloor$, $\gamma = \left\lfloor \frac{n_B - \lambda m(m-1)}{m} \right\rfloor$, and $\delta = n_B - \lambda m(m-1) - \gamma m$}.
We first assign $\lambda$ COAs, $\gamma$ LSs, and $\delta$ rows of LSs to each block. Then, we exchange the LSs and rows of LSs between different blocks to update the design. The {\color{black}construction} algorithm is as follows.\\

{\bf {\color{black}\noindent Construction {\color{black}1}}}

\begin{enumerate}
\item Set an initial WLP for the optimal block OofA design as $ W'_{\text{opt}} = (999, 999, \cdots, 999) $. Note that the optimal block OofA design does not need to exist physically at this step.

\item Select the first $ k\lambda $ COAs, $C_1, \cdots, C_{k\lambda}$, from the set (\ref{eq:setc}) and arbitrarily assign $\lambda$ of them to each block to form an $m$-component block OofA design with $k$ blocks of size $\lambda m(m-1)$, denoted as $d_A'$. If $\lambda = 0$, $d_A'= \emptyset$ (an empty design). 

\item Select LSs $ L_{k\lambda(m-1)+1}, \cdots, L_{k\lambda(m-1)+{\color{black}\left\lceil \frac{k(\gamma m+\delta)}{m} \right\rceil }} $ from the set (\ref{eq:set})  as the candidates. Note that these LSs are included in $ C_{k\lambda+1}, \cdots, C_{k\lambda+\left\lceil \frac{k(\gamma m+\delta)}{m(m-1)} \right\rceil} $. 

\item Perform the following process $I_1$ times.

\begin{enumerate}
\item Arbitrarily assign the $ k\gamma $ candidate LSs to the $ k $ blocks, each of which includes $ \gamma $ of them, to form an $ m $-component block OofA design with $ k $ blocks of size $ \gamma m $, denoted as $ d_B' $. 
If $\gamma=0$, $d_B'=\emptyset$ (an empty design). 

\item Arbitrarily assign $k\delta$ rows from the candidate LSs, which were not selected in step 4(a), to the $k$ blocks, each of which includes $\delta$ of these rows, to form an $m$-component block OofA design with $k$ blocks of size $\delta$, denoted as $d'_C$. 

\item Row-wise concatenate $d_A'$, $d_B'$, and $d_C'$ according to their blocks to form an $m$-component block OofA design with $k$ blocks of size $n_B = \lambda m(m-1)+\gamma m+\delta$, denoted as $D_0'$.

\item Calculate the WLP of $D_0'$, denoted as $W_0'$. 
\item If $\gamma>0$, perform the following process $I_2$ times.
\begin{enumerate}
\item Arbitrarily exchange two LSs between different blocks in $d_B'$ to form $d_B^*$. 
Row-wise concatenate $d_A'$, $d_B^*$, and $d_C'$ according to their blocks to form an updated design, denoted as $D_{up}'$. 
\item Calculate the WLP of $D_{up}'$, denoted as $W_{up}'$. 
\item If $W'_{up} \ll W'_0$, then let $D_0'=D_{up}'$, $W_0'=W_{up}'$, and $d_B'=d_B^*$. 
\end{enumerate}

\item Perform the following process $I_3$ times.
\begin{enumerate}
\item Arbitrarily exchange two rows between different blocks in $d_C'$ to form $d_C^*$. 
Row-wise concatenate $d_A'$, $d_B'$, and $d_C^*$ according to their blocks to form an updated design, denoted as $D_{up}'$. 
\item Calculate the WLP of $D_{up}'$, denoted as $W_{up}'$. 
\item If $W'_{up} \ll W'_0$, then let $D_0'=D_{up}'$, $W_0'=W_{up}'$, and $d_C'=d_C^*$.  
\end{enumerate}

\item If $W'_{0} \ll W'_{opt}$, then let $D_{opt}'=D_{0}'$ and $W_{opt}'=W_{0}'$. 
\end{enumerate} 

\end{enumerate}
The final design $D_{opt}'$ is reported as the optimal $m$-component block OofA design with $k$ blocks of size $n_B=\lambda m(m-1)+\gamma m+\delta$.\\

\begin{table}
\begin{center}
\caption{Block OofA design for $m=5$, $k=3$, and $n_B=12$}\label{tb:12}
\begin{tabular}{ccc}													
{\color{black}
\begin{tabular}{ccc}															
\begin{tabular}{c@{ }c@{}c@{}c@{}c@{}c@{}c@{ }r}															
\hline															
Run	&	$Z_1$	&	$Z_2$	&	$Z_3$	&	$Z_4$	&	$Z_5$	&	$B$	&	\multicolumn{1}{c}{$y$}	\\
\hline															
1	&	1	&	2	&	3	&	4	&	5	&	1	&	29.803	\\
2	&	2	&	3	&	4	&	5	&	1	&	1	&	32.025	\\
3	&	3	&	4	&	5	&	1	&	2	&	1	&	28.397	\\
4	&	4	&	5	&	1	&	2	&	3	&	1	&	19.681	\\
5	&	5	&	1	&	2	&	3	&	4	&	1	&	30.087	\\
6	&	1	&	3	&	5	&	4	&	2	&	1	&	32.708	\\
7	&	2	&	4	&	1	&	5	&	3	&	1	&	29.162	\\
8	&	3	&	5	&	2	&	1	&	4	&	1	&	23.127	\\
9	&	4	&	1	&	3	&	2	&	5	&	1	&	31.067	\\
10	&	5	&	2	&	4	&	3	&	1	&	1	&	31.260	\\
11	&	5	&	3	&	1	&	4	&	2	&	1	&	31.628	\\
12	&	1	&	3	&	5	&	2	&	4	&	1	&	33.295	\\
\hline															
\end{tabular}															
&															
\begin{tabular}{c@{ }c@{}c@{}c@{}c@{}c@{}c@{ }r}															
\hline															
Run	&	$Z_1$	&	$Z_2$	&	$Z_3$	&	$Z_4$	&	$Z_5$	&	$B$	&	\multicolumn{1}{c}{$y$}	\\
\hline															
13	&	1	&	5	&	4	&	3	&	2	&	2	&	10.635	\\
14	&	2	&	1	&	5	&	4	&	3	&	2	&	20.884	\\
15	&	3	&	2	&	1	&	5	&	4	&	2	&	27.398	\\
16	&	4	&	3	&	2	&	1	&	5	&	2	&	26.031	\\
17	&	5	&	4	&	3	&	2	&	1	&	2	&	16.583	\\
18	&	1	&	4	&	2	&	3	&	5	&	2	&	22.322	\\
19	&	2	&	5	&	3	&	4	&	1	&	2	&	9.485	\\
20	&	3	&	1	&	4	&	5	&	2	&	2	&	23.441	\\
21	&	4	&	2	&	5	&	1	&	3	&	2	&	27.882	\\
22	&	5	&	3	&	1	&	2	&	4	&	2	&	26.407	\\
23	&	3	&	1	&	4	&	2	&	5	&	2	&	21.896	\\
24	&	3	&	5	&	2	&	4	&	1	&	2	&	9.561	\\
\hline															
\end{tabular}															
&															
\begin{tabular}{c@{ }c@{}c@{}c@{}c@{}c@{}c@{ }r}															
\hline															
Run	&	$Z_1$	&	$Z_2$	&	$Z_3$	&	$Z_4$	&	$Z_5$	&	$B$	&	\multicolumn{1}{c}{$y$}	\\
\hline															
25	&	1	&	5	&	4	&	2	&	3	&	3	&	10.271	\\
26	&	2	&	1	&	5	&	3	&	4	&	3	&	15.595	\\
27	&	3	&	2	&	1	&	4	&	5	&	3	&	23.651	\\
28	&	4	&	3	&	2	&	5	&	1	&	3	&	19.517	\\
29	&	5	&	4	&	3	&	1	&	2	&	3	&	16.925	\\
30	&	1	&	2	&	3	&	5	&	4	&	3	&	22.037	\\
31	&	2	&	3	&	4	&	1	&	5	&	3	&	24.162	\\
32	&	3	&	4	&	5	&	2	&	1	&	3	&	15.305	\\
33	&	4	&	5	&	1	&	3	&	2	&	3	&	8.629	\\
34	&	5	&	1	&	2	&	4	&	3	&	3	&	19.740	\\
35	&	5	&	2	&	4	&	1	&	3	&	3	&	22.747	\\
36	&	1	&	4	&	2	&	5	&	3	&	3	&	17.681	\\
\hline															
\end{tabular}															
\end{tabular}															
}												
\end{tabular}													
\end{center}
\end{table}

{\bf Example {\color{black}1}}
{\color{black}We provide an example with $m = 5$, $k = 3$, and $n_B = 12$ to illustrate how to construct block OofA designs from COAs and LSs, assign runs to blocks, perform exchanges, and compare WLPs to identify the optimal design.
For $m = 5$, the set $L$ is given in Appendix Table~\ref{tb:ls}, and the set $C$ is obtained according to Equation~(\ref{eq:setc}).
In the case of $m = 5$ and $n_B = 12$, we obtain $\lambda = \left\lfloor \frac{n_B}{m(m - 1)} \right\rfloor = \left\lfloor \frac{12}{5 \times (5 - 1)} \right\rfloor = 0$, $\gamma = \left\lfloor \frac{n_B - \lambda m (m - 1)}{m} \right\rfloor = \left\lfloor \frac{12 - 0 \times 5 \times (5 - 1)}{5} \right\rfloor = 2$, and $\delta = n_B - \lambda m (m - 1) - \gamma m = 12 - 0 \times 5 \times (5 - 1) - 2 \times 5 = 2$.
The algorithm can be simply divided into the following walk-through process: COAs $\rightarrow$ LSs $\rightarrow$ Block assignment $\rightarrow$ Exchange and WLP comparison.

\begin{itemize}
\item[(1)] {\bf COAs}. Select the first $k\lambda$ COAs from the set $C$ as candidate COAs. In this case, $\lambda = 0$, resulting in $k\lambda = 3 \times 0 = 0$. Thus, no candidate COA is required and $d'_A=\emptyset$.

\item[(2)] {\bf LSs}. Select $L_{k\lambda(m-1)+1}, \cdots, L_{k\lambda(m-1)+\left\lceil \frac{k(\gamma m + \delta)}{m} \right\rceil}$ from the set $L$ as candidate LSs. In this case, $k\lambda(m-1) = 3 \times 0 \times (5 - 1) = 0$ and $\left\lceil \frac{k(\gamma m + \delta)}{m} \right\rceil=\left\lceil \frac{3\times(2 \times 5 + 2)}{5} \right\rceil = 8$. Thus, select $L_1, \cdots, L_8$ from the set $L$ as candidate LSs.

\item[(3)] {\bf Block assignment}. 
Since $\gamma = 2 > 0$, arbitrarily assign $k\gamma = 3 \times 2 = 6$ candidate LSs to $k = 3$ blocks to form $d’_B$. Each block of $d’_B$ contains $\gamma = 2$ LSs. Since $\delta = 2 > 0$, arbitrarily assign $k\delta = 3 \times 2 = 6$ rows from the remaining $8 - 6 = 2$ candidate LSs to $k = 3$ blocks to form $d’_C$. Each block of $d’_C$ contains $\delta = 2$ rows. Then, row-wise concatenate $d’_A$, $d’_B$, and $d’_C$ to form $D’_0$.

\item[(4)] {\bf Exchange and WLP comparison}. 
Since $\gamma = 2 > 0$, arbitrarily exchange two LSs between blocks in $d_B'$ to form the updated design $D'_{up}$. Compare $W'_0$ (the WLP of $D'_0$) and $W'_{up}$ (the WLP of $D'_{up}$). If $W'_{up} \ll W'_0$, then replace $D'_0$ with $D'_{up}$.
Since $\delta = 2 > 0$, arbitrarily exchange two rows between blocks in $d_C'$ to form the updated design $D'_{up}$. If $W'_{up} \ll W'_0$, then replace $D'_0$ with $D'_{up}$.
\end{itemize}
}

\noindent {\color{black}Table~\ref{tb:12} presents the resulting block OofA design obtained after executing the algorithm with $I_1 = 500$, $I_2 = 50$, and $I_3 = 50$ iterations. The algorithm} selects $L_1$ and $L_6$ for runs 1-10 in block 1, $L_4$ and $L_7$ for runs 13-22 in block 2, and $L_8$ and $L_5$ for runs 25-34 in block 3. Additionally, row 5 of $L_3$ and row 1 of $L_2$ are used for runs 11-12 in block 1, row 3 of $L_3$ and row 3 of $L_2$ for runs 23-24 in block 2, and row 5 of $L_2$ and row 1 of $L_3$ for runs 35-36 in block 3.
{\color{black}Appendix Table~\ref{tb:27}} shows {\color{black}another} block OofA design for $m=5$, $k=2$, and $n_B=27$, corresponding to the case where $\lambda = 1$, $\gamma = 1$, and $\delta = 2$.
The candidates for constructing the design are COAs $C_1$ and $C_2$, and LSs $L_9, \cdots, L_{12}$ (included in $C_3$).
First, in step 2, $C_1$ and $C_2$ are assigned to runs 1-20 in block 1 and runs 28-47 in block 2, respectively.
Then, after performing step 4 $I_1 = 500$ times, step 4(e) $I_2 = 50$ times, and step 4(f) $I_3 = 50$ times, the algorithm finally selects $L_{9}$ for runs 21-25 in block 1, $L_{11}$ for runs 48-52 in block 2, rows 2 and 4 of $L_{10}$ for runs 26-27 in block 1, and rows 5 and 3 of $L_{10}$ for runs 53-54 in block 2.  
Their word length patterns can be found in Table~\ref{tb:wlp}. 
It shows that the two designs achieve small $w_l^P$ and $w_l^B$ values that are close to, or equal to, those of the full block OofA designs.

\begin{table}
\begin{center}
\caption{WLPs of the full and construed block OofA designs}\label{tb:wlp}
{\color{black}
\begin{tabular}{ccccccccccc}																					
\hline																					
$k$	&	$n_B$	&	Design	&	$w_1^P$	&	$w_1^B$	&	$w_2^P$	&	$w_2^B$	&	$w_3^P$	&	$w_3^B$	&	$w_4^P$	&	$w_4^B$	\\
\hline																					
3	&	120	&	Full	&	0	&	0	&	0.625	&	0	&	0	&	0	&	1.408	&	0	\\
	&	20	&	Table \ref{tb:20}	&	0	&	0	&	0.625	&	0	&	0	&	0	&	1.527	&	0.476	\\
	&	15	&	Table \ref{tb:15}	&	0	&	0	&	0.633	&	0.061	&	0.110	&	1.517	&	1.600	&	1.077	\\
	&	12	&	Table \ref{tb:12}	&	0	&	0	&	0.687	&	0.317	&	0	&	1.901	&	1.954	&	4.393	\\
\hline																					
2	&	120	&	Full	&	0	&	0	&	0.625	&	0	&	0	&	0	&	1.408	&	0	\\
	&	40	&	Table \ref{tb:40}	&	0	&	0	&	0.625	&	0	&	0	&	0	&	1.468	&	0.179	\\
	&	27	&	Table \ref{tb:27}	&	0.002	&	0.005	&	0.633	&	0.042	&	0.086	&	0.199	&	1.564	&	0.562	\\
	&	25	&	Table \ref{tb:25}	&	0	&	0	&	0.625	&	0.025	&	0.179	&	0.179	&	1.546	&	0.579	\\
\hline																					
\end{tabular}																											
}																						
\end{center}
\end{table}

\subsection{Block size $n_B$ is a multiple of $m$}
{\color{black}When $n_B$ is a multiple of $m$, i.e., }$n_B = rm$, where $r = 1, 2, \cdots, \left\lfloor \frac{(m-1)!}{k} \right\rfloor${\color{black}, %
we} divide this situation into the following two cases. 
The first case, discussed in Section~\ref{se:r2}, is when $ r $ is not a multiple of $ (m-1) $.
The second case, discussed in Section~\ref{se:r1}, is when $ r $ is a multiple of $ (m-1) $.

\subsubsection{Block Size $ n_B = rm $ when $ r $ is not a multiple of $ (m-1) $}\label{se:r2}

In this case, $r$ can be rewritten as $r=\lambda(m-1)+\gamma$, where $\lambda=0,1,\cdots,\left\lfloor \frac{(m-2)!}{k} \right\rfloor-1$ and $\gamma=1,\cdots,m-2$. 
Thus, the block size can be expressed as $n_B = \lambda m(m-1)+\gamma m$. 
Since a COA is an $m(m-1) \times m$ array and an LS is an $m \times m$ array, we first assign $\lambda$ COAs and $\gamma$ LSs from the candidate sets to each block and then exchange the LSs between different blocks to update the design.
{\color{black}The construction algorithm is a special case of Construction~1 obtained by setting $\gamma > 0$ and $\delta = 0$.}
The final design $ D_{opt}' $ is reported as the optimal $ m $-component block OofA design with $ k $ blocks of size $n_B = \lambda m(m-1)+\gamma m$. \\

{\bf Example 2 } {\color{black}Appendix Table~\ref{tb:15}} shows a block OofA design for $m=5$, $k=3$, and $n_B=15$. 
This corresponds to the case where $\lambda = 0$ and $\gamma = 3$. 
Thus, $d_A' = \emptyset$ in step 2, and LSs $L_1, \cdots, L_{12}$ (included in $C_1, \cdots, C_3$) are selected as candidates in step 3.
After performing step 4 $I_1 = 500$ times and step 4(d) $I_2 = 50$ times, the algorithm finally selects $L_1$, $L_4$, and $L_3$ for block 1; $L_7$, $L_8$, and $L_9$ for block 2; and $L_5$, $L_2$, and $L_6$ for block 3.
{\color{black}Appendix Table~\ref{tb:25}} shows a block OofA design for $m=5$, $k=2$, and $n_B=25$. 
This corresponds to the case where $\lambda = 1$ and $\gamma = 1$. 
The candidates for constructing the design are COAs $C_1$ and $C_2$, and LSs $L_9, \cdots, L_{12}$ (included in $C_3$). 
First, in step 2, $C_1$ and $C_2$ are assigned to runs 1-20 in block 1 and runs 26-45 in block 2, respectively. 
Then, after performing step 4 $I_1 = 500$ times and step 4(d) $I_2 = 50$ times, the algorithm finally selects $L_{10}$ and $L_{9}$ for runs 21-25 in block 1 and runs 46-50 in block 2, respectively.
The WLPs of the two block OofA designs can be found in Table~\ref{tb:wlp}. 
It shows that both constructed designs achieve $w_1^P = w_3^P =  w_1^B = 0$ and have small $w_2^P$, $w_2^B$, $w_3^B$, and $w_4^B$, which are close to those of the full block OofA designs.

\subsubsection{Block Size $ n_B = rm $ when $ r $ is a multiple of $ (m-1) $}\label{se:r1}

In this case, $r$ can be rewritten as $r=\lambda(m-1)$, where $\lambda=1,2,\cdots,\left\lfloor \frac{(m-2)!}{k} \right\rfloor$. 
Thus, the block size can be expressed as $n_B = \lambda m(m-1)$. 
Since a COA is an $m(m-1) \times m$ array, we can construct such a block OofA design by assigning $\lambda$ COAs to each block. 
{\color{black}The construction algorithm is a special case of Construction 1 with $\lambda > 0$, $\gamma = 0$, and $\delta=0$.}
{\color{black}The final design $ D_{opt}' $ is reported as the optimal $ m $-component block OofA design with $ k $ blocks of size $n_B = \lambda m(m-1)$.}
The algorithm constructs the block OofA design directly from the COAs without any exchange and search process.
It ensures that the resulting design has ${\color{black}w_1^B} = w_2^B = w_3^B = 0 $, $ w_1^P = w_3^P = 0 $, and the smallest $ w_2^P $, which equals that of the full block OofA design.\\

{\bf Example {\color{black}3}} 
When $m=5$, $k=3$, and $n_B=20$ (i.e., $\lambda=1$), we select $k\lambda=3$ candidate COAs, $C_1, C_2, C_3$, and assign $C_1$ to block 1, $C_2$ to block 2, and $C_3$ to block 3, forming a block OofA design with 3 blocks of size 20, as shown in {\color{black}Appendix Table~\ref{tb:20}}. Similarly, for $k=2$ and $n_B=40$ (i.e., $\lambda=2$), we select $k\lambda=4$ candidate COAs, $C_1, \cdots, C_4$, and assign $C_1$ and $C_2$ to block 1 and $C_3$ and $C_4$ to block 2, forming a block OofA design with 2 blocks of size 40, as shown in {\color{black}Appendix Table~\ref{tb:40}}. 
The word length patterns for $w^P_1,\cdots,w^P_4$ and ${\color{black}w^B_1},\cdots,w^B_4$ of the two block OofA designs can be found in Table~\ref{tb:wlp}. 
For benchmark, we also include those of the full block OofA designs with $k=2$ and $k=3$ in the table. 
The results in Table~\ref{tb:wlp} verify that the constructed block OofA designs achieve $w_l^B =  0$ for $l\leq 3$, $w_1^P = w_3^P = 0$, and the smallest $w_2^P$, which equals that of the full block OofA design.\\

Note that the block OofA design for the 5-drug experiment described by Mee (2020) is a specific instance that can be constructed using this algorithm with $m = 5$, $k = 2$, and $n_B = 20$. 
This approach is efficient and straightforward. 
When $n_B = rm$, where $r$ is a multiple of $(m-1)$, block OofA designs can be directly obtained using {\color{black}Construction}~1, with no additional computations required.

{\color{black}A practical cookbook for the required processes (COA: COA assignment; LSex: LS exchange; ROWex: row exchange) and the suggested default settings for the iteration parameters $(I_1, I_2, I_3)$ is provided in Appendix Table~\ref{tb:cook}.
{\color{black}When $m$ is not a prime power, a Galois field of order $m$ does not exist. In such cases, the starting designs can be generated by randomly selecting $k n_B$ points from the full $m$-component OofA design and randomly assigning them to $k$ blocks of size $n_B$.
Then, the Fedorov's point-exchange algorithm (Fedorov, 1972) can be applied to obtain optimal block OofA designs. 
A more systematic construction method for this situation will be considered in future work.}


{\color{black}
\begin{table}
\begin{center}
\caption{Five-drug experiment by randomly assigning instances from three batchs}\label{tb:unbk}
{\color{black}
\begin{tabular}{cc}																	
\begin{tabular}{cc@{ }c@{ }c@{ }c@{ }c@{ }r}																	
\hline																	
Run	&	$Z_1$	&	$Z_2$	&	$Z_3$	&	$Z_4$	&	$Z_5$	&		\multicolumn{1}{c}{$y$}	\\			
\hline																	
1	&	1	&	2	&	3	&	4	&	5	&	$	30.873	^	1	$	\\
2	&	2	&	3	&	4	&	5	&	1	&	$	19.370	^	3	$	\\
3	&	3	&	4	&	5	&	1	&	2	&	$	17.879	^	3	$	\\
4	&	4	&	5	&	1	&	2	&	3	&	$	19.938	^	1	$	\\
5	&	5	&	1	&	2	&	3	&	4	&	$	23.081	^	2	$	\\
6	&	1	&	3	&	5	&	4	&	2	&	$	21.962	^	3	$	\\
7	&	2	&	4	&	1	&	5	&	3	&	$	21.778	^	2	$	\\
8	&	3	&	5	&	2	&	1	&	4	&	$	13.716	^	3	$	\\
9	&	4	&	1	&	3	&	2	&	5	&	$	21.600	^	2	$	\\
10	&	5	&	2	&	4	&	3	&	1	&	$	23.452	^	2	$	\\
11	&	5	&	3	&	1	&	4	&	2	&	$	31.106	^	1	$	\\
12	&	1	&	3	&	5	&	2	&	4	&	$	21.519	^	3	$	\\
13	&	1	&	5	&	4	&	3	&	2	&	$	8.605	^	3	$	\\
14	&	2	&	1	&	5	&	4	&	3	&	$	29.623	^	1	$	\\
15	&	3	&	2	&	1	&	5	&	4	&	$	26.617	^	2	$	\\
16	&	4	&	3	&	2	&	1	&	5	&	$	27.118	^	2	$	\\
17	&	5	&	4	&	3	&	2	&	1	&	$	17.648	^	2	$	\\
18	&	1	&	4	&	2	&	3	&	5	&	$	30.525	^	1	$	\\
\hline																	
\end{tabular}																	
&																	
\begin{tabular}{cc@{ }c@{ }c@{ }c@{ }c@{ }r}																	
\hline																	
Run	&	$Z_1$	&	$Z_2$	&	$Z_3$	&	$Z_4$	&	$Z_5$	&		\multicolumn{1}{c}{$y$}	\\			
\hline																	
19	&	2	&	5	&	3	&	4	&	1	&	$	18.024	^	1	$	\\
20	&	3	&	1	&	4	&	5	&	2	&	$	20.871	^	3	$	\\
21	&	4	&	2	&	5	&	1	&	3	&	$	33.739	^	1	$	\\
22	&	5	&	3	&	1	&	2	&	4	&	$	26.543	^	2	$	\\
23	&	3	&	1	&	4	&	2	&	5	&	$	26.742	^	1	$	\\
24	&	3	&	5	&	2	&	4	&	1	&	$	5.319	^	2	$	\\
25	&	1	&	5	&	4	&	2	&	3	&	$	10.767	^	3	$	\\
26	&	2	&	1	&	5	&	3	&	4	&	$	29.046	^	1	$	\\
27	&	3	&	2	&	1	&	4	&	5	&	$	32.916	^	1	$	\\
28	&	4	&	3	&	2	&	5	&	1	&	$	24.632	^	2	$	\\
29	&	5	&	4	&	3	&	1	&	2	&	$	16.860	^	3	$	\\
30	&	1	&	2	&	3	&	5	&	4	&	$	21.139	^	3	$	\\
31	&	2	&	3	&	4	&	1	&	5	&	$	27.633	^	2	$	\\
32	&	3	&	4	&	5	&	2	&	1	&	$	14.690	^	3	$	\\
33	&	4	&	5	&	1	&	3	&	2	&	$	11.618	^	2	$	\\
34	&	5	&	1	&	2	&	4	&	3	&	$	31.630	^	1	$	\\
35	&	5	&	2	&	4	&	1	&	3	&	$	33.366	^	1	$	\\
36	&	1	&	4	&	2	&	5	&	3	&	$	19.057	^	3	$	\\
\hline																	
\end{tabular}																	
\end{tabular}
}																	
\end{center}
\end{table}

\begin{table}
\begin{center}
\caption{Forward regression for the experiment with the unblock OofA design in Table~\ref{tb:unbk}}\label{tb:reg_unbk}
{\color{black}
\begin{tabular}{lrrrr@{ }l}											
\hline											
	&	Estimate	&	Std. Error	&	t value	&	Pr($>|$t$|$)	&		\\
(Intercept)	&	22.4438	&	0.7232	&	31.034	&	$<$ 2.00e-16	&	***	\\
$Z_2^l$	&	-4.3377	&	0.7998	&	-5.423	&	5.80e-06	&	***	\\
$Z_2^q$	&	-2.5307	&	0.7189	&	-3.52	&	0.00132	&	**	\\
$Z_5^l$	&	1.9279	&	0.7998	&	2.41	&	0.02186	&	*	\\
\hline											
\end{tabular}											
}
\end{center}
\end{table}

\begin{table}
\begin{center}
\caption{Forward regression for the experiment with the block OofA design in Table~\ref{tb:12}}\label{tb:reg_12}
{\color{black}
\begin{tabular}{lrrrr@{ }l}											
\hline											
	&	Estimate	&	Std. Error	&	t value	&	Pr($>|$t$|$)	&		\\
(Intercept)	&	23.0018	&	0.1915	&	120.107	&	$<$ 2.00e-16	&	***	\\
$B^l$	&	-4.3883	&	0.1669	&	-26.287	&	$<$ 2.00e-16	&	***	\\
$Z_2^l$	&	-3.2385	&	0.1792	&	-18.075	&	$<$ 2.00e-16	&	***	\\
$Z_2^q$	&	-3.1034	&	0.1864	&	-16.652	&	1.00e-15	&	***	\\
$B^q$	&	1.0130	&	0.1668	&	6.072	&	1.75e-06	&	***	\\
$Z_5^l$	&	1.0476	&	0.1792	&	5.847	&	3.17e-06	&	***	\\
$Z_2^lZ_5^l$	&	1.4687	&	0.2291	&	6.410	&	7.24e-07	&	***	\\
$Z_1^lZ_5^l$	&	0.9691	&	0.1965	&	4.932	&	3.65e-05	&	***	\\
$Z_3^lZ_4^l$	&	-0.6595	&	0.1993	&	-3.309	&	0.00266	&	**	\\
\hline											
\end{tabular}												
}			
\end{center}
\end{table}

\section{Five-drug experiment}\label{se:5dru}
In real-world applications, OofA experiments are commonly used to study drug combination therapy for cancer treatment, viral infection eradication, and super bacteria inhibition, where the sequence of drug administration is believed to play a key role. Such experiments are often conducted across multiple groups or batches (see Mee, 2020; Wang et al., 2020; Yang et al., 2021; and Stokes and Xu, 2022).  
To demonstrate the importance of block OofA designs in mitigating batch-by-batch variation, we present the following {\color{black}example}. 

Consider a combination therapy involving five drugs{\color{black}, denoted as $\text{drug}1,\cdots,\text{drug}5$,} with an underlying true model $\mu = 23.13 + 0.26Z_1^l - 3.19Z_2^l + 1.3Z_5^l - 3.21Z_2^q + 1.05Z_1^lZ_5^l + 1.82Z_2^lZ_5^l$.  
By evaluating all $5! = 120$ possible five-drug sequences, the highest value of $\mu$ (where a larger value is better) is achieved when the sequence is either {\color{black}$\text{drug}4 \rightarrow \text{drug}3 \rightarrow \text{drug}2 \rightarrow \text{drug}1 \rightarrow \text{drug}5$ or $\text{drug}3 \rightarrow \text{drug}4 \rightarrow \text{drug}2 \rightarrow \text{drug}1 \rightarrow \text{drug}5$.}
These two sequences represent the optimal drug order for the combination therapy. 

A researcher aims to conduct an experiment to determine the optimal drug sequence for the therapy, involving three groups of patients, labeled $B = 1, 2, 3$. The patients from each group introduce batch effects, $- 4.08B^l + 1.2B^q$, into the response. 
However, the experimenter is unaware of the batch-to-batch variation and ignores it by randomly assigning patients from the three groups to the 36 runs of an OofA design. 
The design and its responses, obtained based on the true model, batch effects, and experimental random error $ \epsilon \sim N(0,1) $, are shown in Table~\ref{tb:unbk}, where the superscript of the $y$ value indicates which group of patients the run was conducted with.
Note that the 36 runs of the OofA designs in Table~\ref{tb:unbk} are identical to those in Table~\ref{tb:12}. 
The researcher performs forward regression analysis using the full second-order position model with $ \alpha = 0.05 $. The results of this analysis are presented in Table~\ref{tb:reg_unbk}. 
It identifies that only the effects of $ Z_2^l $, $ Z_2^q $, and $ Z_5^l $ are significant, with the fitted model $\hat y = 22.4438 - 4.3377Z_2^l + 1.9279Z_5^l - 2.5307Z_2^q$.
Based on the fitted model, the highest predicted response {\color{black}(29.750)} occurs when the sequence is either 
{\color{black}$\text{drug}1 \rightarrow \text{drug}2 \rightarrow \text{drug}3 \rightarrow \text{drug}4 \rightarrow \text{drug}5$,  
$\text{drug}3 \rightarrow \text{drug}2 \rightarrow \text{drug}1 \rightarrow \text{drug}4 \rightarrow \text{drug}5$,  
$\text{drug}1 \rightarrow \text{drug}2 \rightarrow \text{drug}4 \rightarrow \text{drug}3 \rightarrow \text{drug}5$, 
$\text{drug}4 \rightarrow \text{drug}2 \rightarrow \text{drug}3 \rightarrow \text{drug}1 \rightarrow \text{drug}5$, 
$\text{drug}4 \rightarrow \text{drug}2 \rightarrow \text{drug}1 \rightarrow \text{drug}3 \rightarrow \text{drug}5$, 
or $\text{drug}3 \rightarrow \text{drug}2 \rightarrow \text{drug}4 \rightarrow \text{drug}1 \rightarrow \text{drug}5$.}   
None of these sequences match the optimal drug sequences specified by the true model. 
{\color{black}The true responses corresponding to these six sequences are 24.855, 27.322, 24.855, 28.556, 27.322, and 28.556, respectively. Compared with the true responses, the prediction errors of the six sequences are $4.895$, $2.427$, $4.895$, $1.193$, $2.427$, and $1.193$, respectively.}
This result demonstrates that by ignoring batch-to-batch variation, the experiment fails to identify the true optimal drug sequence for the combination therapy.

Now, another researcher, aware of the batch-to-batch variation, conducts the experiment using the block OofA design generated according to the algorithm in Section~\ref{se:al}.
The design and its responses, obtained based on the true model, batch effects, and experimental random error $ \epsilon \sim N(0,1) $, are given in Table~\ref{tb:12}. 
The result of the forward regression analysis is presented in Table~\ref{tb:reg_12}. It gives the fitted model $\hat y = 23.0018 - 4.3883B^l + 1.0130B^q - 3.2385Z_2^l + 1.0476Z_5^l - 3.1034Z_2^q + 0.9691Z_1^lZ_5^l + 1.4687Z_2^lZ_5^l - 0.6595Z_3^lZ_4^l$. 
Among the six identified active position terms, five are included in the true model, except for $Z_3^lZ_4^l$.
Although $Z_3^lZ_4^l$ is falsely identified as active, its $p$-value is relatively large (0.003) compared to those of the five active position terms ($\leq 3.65 \times 10^{-5}$). 
Based on the fitted model, for a given batch, the highest predicted response occurs when the sequence is either  
{\color{black}$\text{drug}4 \rightarrow \text{drug}3 \rightarrow \text{drug}2 \rightarrow \text{drug}1 \rightarrow \text{drug}5$  
or  
$\text{drug}3 \rightarrow \text{drug}4 \rightarrow \text{drug}2 \rightarrow \text{drug}1 \rightarrow \text{drug}5$},  
which match the optimal sequences specified by the true model.

{\color{black}
A short robustness check using forward regression with AIC criterion is conducted.
For the unblock OofA design, the selection steps are listed in Appendix Table~\ref{tb:aic_unbk}.
The results show that the AIC decreases rapidly from step 1 to 4 (from 144.0 to 109.4), while the rate of decrease slows from step 4 to 5 (from 109.4 to 108.7).
This suggests that the best model is the one at step 4, which is identical to the model selected by forward regression using $\alpha = 0.05$.
For the block OofA design, the model selection steps are listed in Appendix Table~\ref{tb:aic_bk}.
Similarly, the AIC decreases rapidly from forward step 1 to 9 (from 144.0 to 5.6), and the reduction slows from step 9 to 10 (from 5.6 to 3.8).
This suggests that the best model is the one at step 9, which is also identical to the model selected by forward regression using $\alpha = 0.05$.
This robustness check confirms that the qualitative conclusions do not depend on the model selection scheme.
}

This example demonstrates a real-world application of the block OofA design in drug combination therapy when there is variance between batches. It shows that performing experiments using the proposed block OofA design can account for batch-by-batch variation and effectively identify significant position effects of the components.

\begin{figure}
\begin{subfigure}{.5\textwidth}
  \centering
  \includegraphics[width=.8\linewidth]{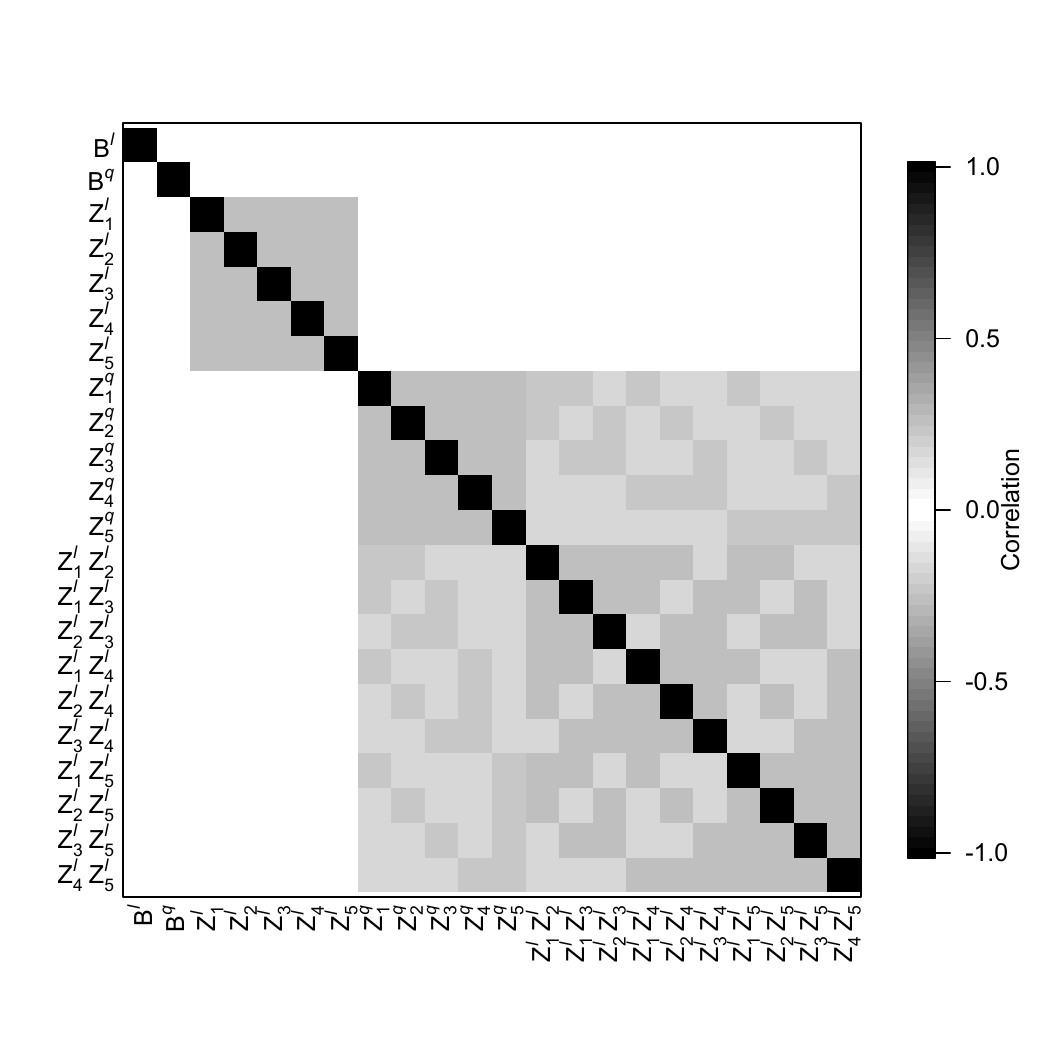}
  \caption{$n_B=120$ (Full design)}
  \label{fi:3x120}
\end{subfigure}%
\begin{subfigure}{.5\textwidth}
  \centering
  \includegraphics[width=.8\linewidth]{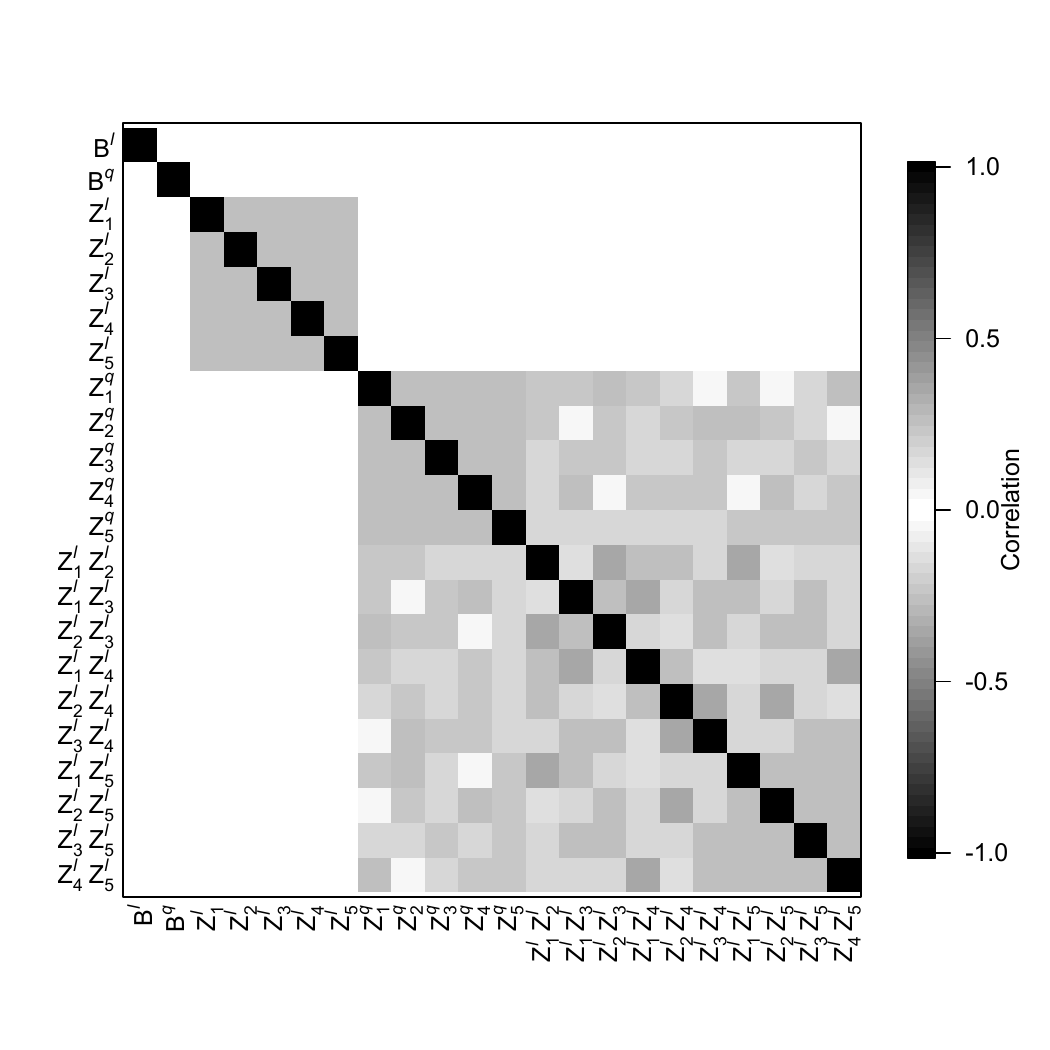}
  \caption{$n_B=20$ (Table \ref{tb:20})}
  \label{fi:20}
\end{subfigure}%

\begin{subfigure}{.5\textwidth}
  \centering
  \includegraphics[width=.8\linewidth]{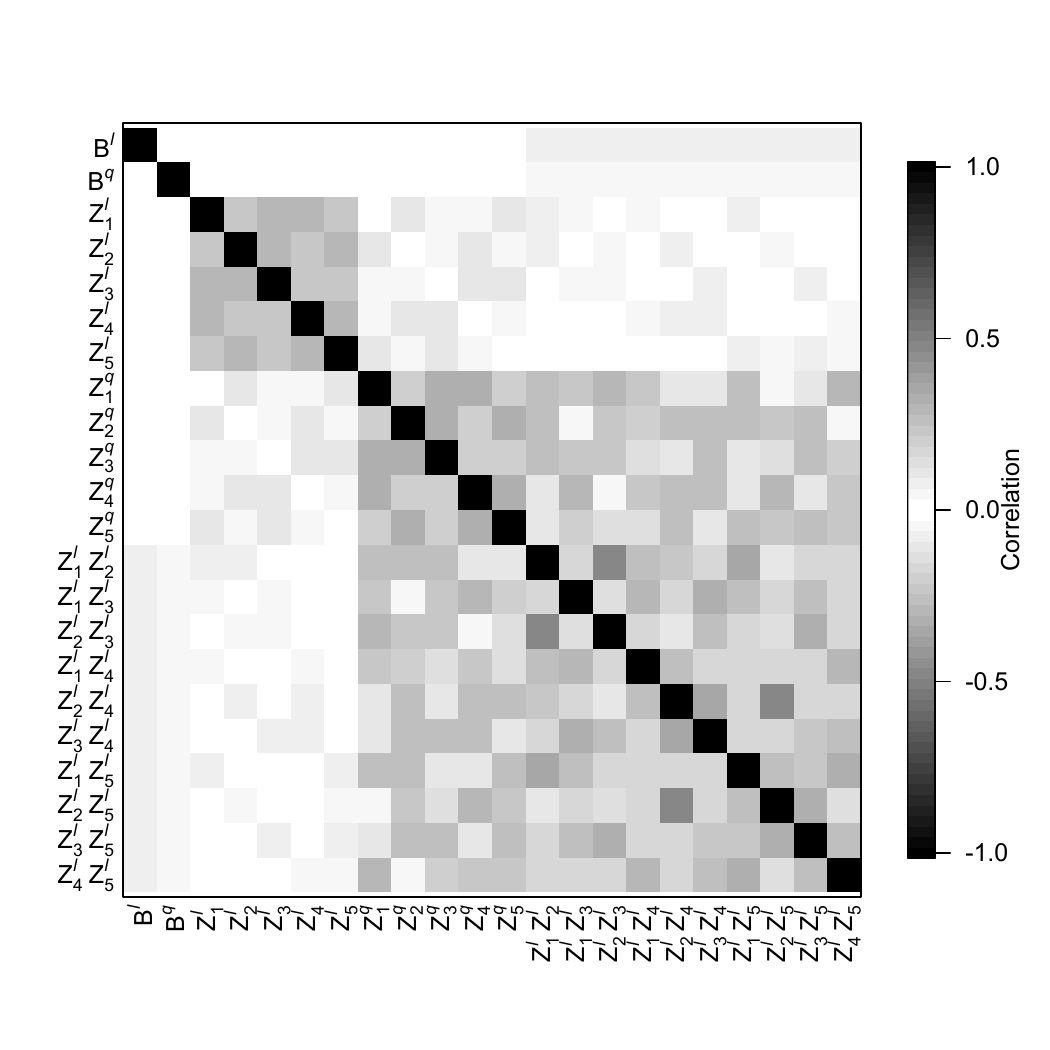}
  \caption{$n_B=15$ (Table \ref{tb:15})}
  \label{fi:15}
\end{subfigure}%
\begin{subfigure}{.5\textwidth}
  \centering
  \includegraphics[width=.8\linewidth]{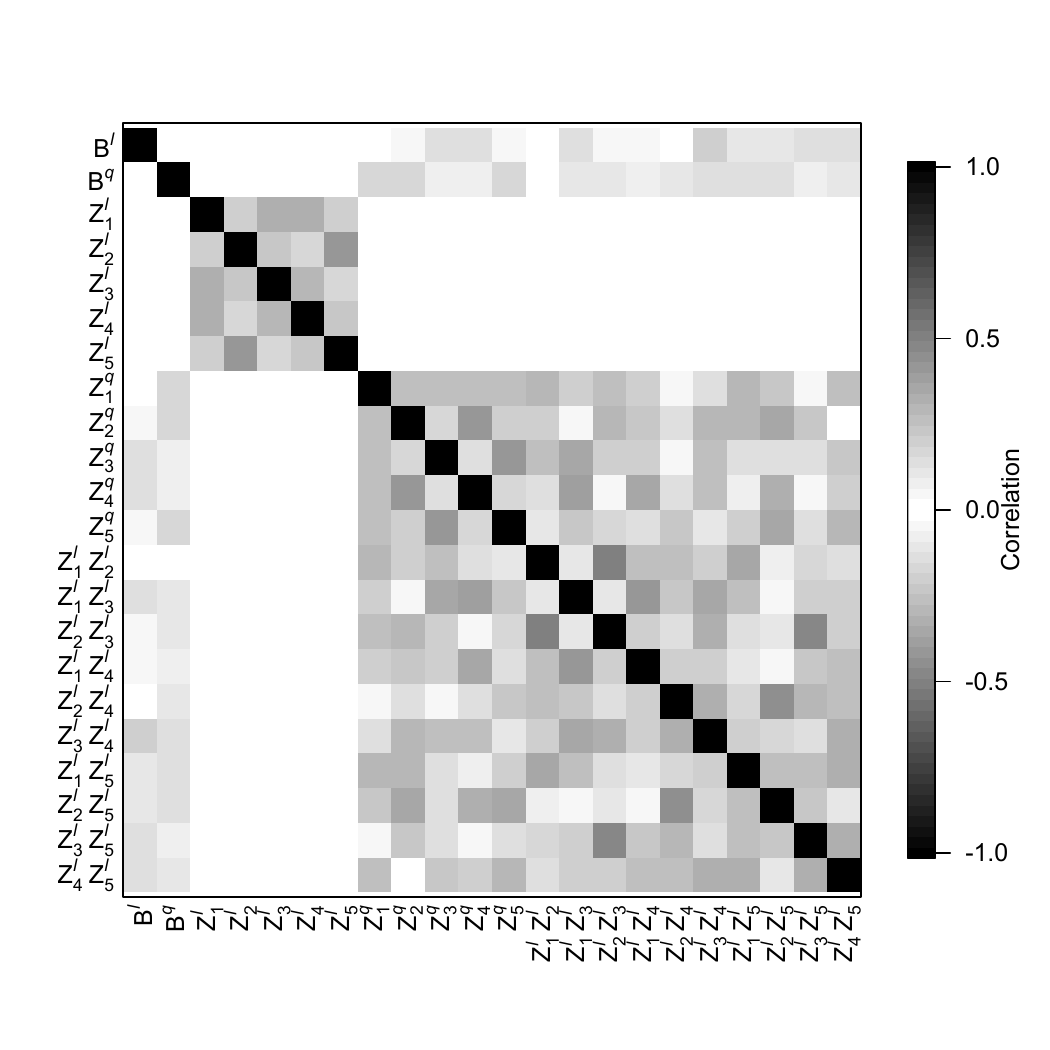}
  \caption{$n_B=12$ (Table \ref{tb:12})}
  \label{fi:12}
\end{subfigure}%
\caption{Correlation plots of block OofA designs for $m=5,k=3$}
\label{fi:k3}
\end{figure}

\section{Simulation study and evaluation}\label{se:sim}
To evaluate the performance of the block OofA designs constructed based on the proposed criterion and algorithms, we examine the correlation between two columns in the full second-order block-position model matrix of these designs. Additionally, we conduct a simulation study to assess their power and type I error rate for identifying active effects.

Figure~\ref{fi:k3} (for $k=3$) and {\color{black}Appendix Figure~\ref{fi:k2}} (for $k=2$) show the correlation plots of the block OofA designs constructed in Section~\ref{se:al}, where $B^l$ and $B^q$ represent the linear and quadratic block effects, respectively; $Z_j^l$ and $Z_j^q$ for $j=1, \cdots, 5$ represent the linear and quadratic position effects, respectively, of factor $Z_j$; and $Z_i^lZ_j^l$ for $1 \leq i < j \leq m$ represents the interaction of the linear effects of $Z_i$ and $Z_j$.
For the benchmark, Figure~\ref{fi:3x120} presents the correlation plot of the full block OofA design for $k=3$. 
It shows that, for the full block design, the block effects are uncorrelated with all of the position effects and interactions. Additionally, linear position effects are uncorrelated with quadratic position effects and the interaction of linear effects of two position factors.
Figure~\ref{fi:20} is the correlation plot of the block OofA design with $n_B=20$ in Table~\ref{tb:20}. 
This design achieves the same uncorrelated structure as the full block OofA design shown in Figure~\ref{fi:3x120}. 
Figure~\ref{fi:15} is the correlation plot of the block OofA design with $n_B=15$ in {\color{black}Appendix Table~\ref{tb:15}}, showing that the block effects are uncorrelated with the linear position effects. Additionally, it demonstrates that the proposed method effectively reduces the correlations between the block effects and the interactions of linear position effects, as well as between the linear position effects and their interactions.
Figure~\ref{fi:12} is the correlation plot of the block OofA design with $n_B=12$ in Table~\ref{tb:12}. In this design, the linear position effects and the interaction of linear position effects are also uncorrelated.
For the designs with $k=2$, their correlation plots are shown in {\color{black}Appendix Figure~\ref{fi:k2}}. 
They exhibit similar correlation patterns to those in Figure~\ref{fi:k3} for $k=3$.

\begin{table}
\begin{center}
\caption{{\color{black}Power ($PW$), type I error rate ($TY1$), and response difference between the true and estimated optimal orders ($DIF$)} of the designs in the simulation study}\label{tb:pt}
{\color{black}
\begin{tabular}{cccccccccc}																			
\hline																			
$k$	&	$n_B$	&	Design	&		&	$p=1$	&	$p=2$	&	$p=3$	&	$p=4$	&	$p=5$	&	$p=6$	\\
\hline																			
3	&	120	&	Full design	&	$PW$	&	1.000	&	1.000	&	0.962	&	0.907	&	0.924	&	0.894	\\
	&		&		&	$TY1$	&	0.041	&	0.040	&	0.053	&	0.068	&	0.065	&	0.088	\\
	&		&		&	$DIF$	&	0.140	&	0.133	&	0.126	&	0.129	&	0.135	&	0.134	\\
	&	20	&	Table \ref{tb:20}	&	$PW$	&	1.000	&	1.000	&	0.946	&	0.887	&	0.902	&	0.861	\\
	&		&		&	$TY1$	&	0.044	&	0.043	&	0.058	&	0.070	&	0.073	&	0.095	\\
	&		&		&	$DIF$	&	0.356	&	0.334	&	0.308	&	0.308	&	0.339	&	0.336	\\
	&	15	&	Table \ref{tb:15}	&	$PW$	&	1.000	&	1.000	&	0.943	&	0.871	&	0.892	&	0.849	\\
	&		&		&	$TY1$	&	0.046	&	0.044	&	0.059	&	0.073	&	0.075	&	0.100	\\
	&		&		&	$DIF$	&	0.412	&	0.388	&	0.349	&	0.372	&	0.367	&	0.387	\\
	&	12	&	Table \ref{tb:12}	&	$PW$	&	1.000	&	1.000	&	0.938	&	0.876	&	0.882	&	0.846	\\
	&		&		&	$TY1$	&	0.051	&	0.049	&	0.064	&	0.075	&	0.085	&	0.103	\\
	&		&		&	$DIF$	&	0.472	&	0.458	&	0.412	&	0.399	&	0.436	&	0.449	\\
\hline																			
2	&	120	&	Full design	&	$PW$	&	1.000	&	1.000	&	0.956	&	0.887	&	0.904	&	0.878	\\
	&		&		&	$TY1$	&	0.040	&	0.041	&	0.050	&	0.066	&	0.068	&	0.086	\\
	&		&		&	$DIF$	&	0.175	&	0.165	&	0.154	&	0.152	&	0.165	&	0.169	\\
	&	40	&	Table \ref{tb:40}	&	$PW$	&	1.000	&	1.000	&	0.939	&	0.865	&	0.887	&	0.849	\\
	&		&		&	$TY1$	&	0.043	&	0.041	&	0.057	&	0.069	&	0.071	&	0.092	\\
	&		&		&	$DIF$	&	0.309	&	0.291	&	0.265	&	0.276	&	0.285	&	0.296	\\
	&	27	&	Table \ref{tb:27}	&	$PW$	&	1.000	&	1.000	&	0.932	&	0.857	&	0.879	&	0.839	\\
	&		&		&	$TY1$	&	0.044	&	0.043	&	0.058	&	0.072	&	0.075	&	0.096	\\
	&		&		&	$DIF$	&	0.376	&	0.353	&	0.317	&	0.336	&	0.340	&	0.359	\\
	&	25	&	Table \ref{tb:25}	&	$PW$	&	1.000	&	1.000	&	0.926	&	0.858	&	0.873	&	0.840	\\
	&		&		&	$TY1$	&	0.045	&	0.044	&	0.059	&	0.072	&	0.076	&	0.100	\\
	&		&		&	$DIF$	&	0.389	&	0.371	&	0.349	&	0.347	&	0.362	&	0.371	\\
\hline																			
\end{tabular}																			}																									
\end{center}
\end{table}

To examine the performance of the constructed block OofA designs in Section~\ref{se:al}, we perform the following simulation study to evaluate their power and type I error rate with the full second-order block-position model defined in Section~\ref{se:bpb}.

\begin{enumerate}

\item {\color{black}Following the strong heredity principle (Wu and Hamada, 2011), we randomly select $p$ position effects to be active by first randomly selecting $p_1$ linear effects, and then, among these effects, randomly selecting $p_2$ pairs for interaction effects and $p_3$ quadratic effects such that $p_1 + p_2 + p_3 = p$.
Their} effect sizes are randomly drawn from a uniform distribution $\pm U[2\sigma_\epsilon, 4\sigma_\epsilon]$, where $\sigma_\epsilon=1$ is the standard deviation of the random error $\epsilon$. The remaining position effects are set to be inactive with zero effects.

\item All block effects are set to be active with effect size drawn from a uniform distribution $ \pm U[2\sigma_\epsilon, 4\sigma_\epsilon] $.

\item Generate the response vector $\by$ from a multivariate normal distribution $MN(\bX\bbeta, \sigma^2_\epsilon\bI)$, where $\bX$ is the model matrix and $\bbeta$ is the vector comprising the intercept, position effects set from step 1, and block effects set from step 2.

\item Perform forward variable selection in the regression with a significance level $\alpha = 0.05$ to sequentially identify significant block or position effects.

\item Calculate the power (\(pw\)) as 
\begin{equation} 
pw = \frac{N_{TruePositive}}{N_{Act}}, 
\end{equation} 
where \(N_{Act} = (k-1) + p\) is the total number of active block and position effects, and \(N_{TruePositive}\) is the number of active effects identified as significant, and calculate the type I error rate (\(ty1\)) as
\begin{equation} 
ty1 = \frac{N_{FalsePositive}}{N_{Inact}},
\end{equation} 
where \(N_{Inact}\) is the number of inactive position effects, and \(N_{FalsePositive}\) is the number of inactive effects identified as significant.

\item Perform steps 1 to 5 $I=1000$ times to obtain {\color{black}the power $pw_i$, the type I error rate $ty1_i$, the response $y_{i(opt)}$ of the true optimal order, and the response ${\hat y}_{i(opt)}$ of the estimated optimal order} for $i=1, \cdots, I$. The overall power and type I error rate are then evaluated by

\begin{equation} 
PW = \frac{\sum_{i=1}^I pw_i}{I}
\end{equation} 
and
\begin{equation} 
TY1 = \frac{\sum_{i=1}^I ty1_i}{I}.
\end{equation} 
{\color{black}The average difference between the responses of the true optimal order and the estimated optimal order is calculated as
\begin{equation} 
DIF= \frac{\sum_{i=1}^I |{y_{i(opt)}-\hat y_{i(opt)}}|}{I},
\end{equation} 
which quantifies how close the responses of the estimated optimal order are to those of the true optimal order.}
\end{enumerate}

For comparison, we also conduct simulations for full block OofA designs with $k=2$ and $k=3$ as benchmarks. Based on the effect sparsity principle (see Wu {\color{black}and} Hamada, {\color{black}2021}), we consider the number of active position effects $p \leq 6$. The results are listed in Table~\ref{tb:pt}. They show that the powers for all designs reach 1 when the number of active position effects $p = 1$ and $2$. Generally, as $p$ increases and $n_B$ decreases, we observe a decrease in power and an increase in type I error rate.
Compared to full block OofA designs, the constructed designs achieve smaller block sizes without significant loss in power. For instance, when $p=6$, the design with $k=2$ and $n_B=25$ (Table~\ref{tb:25}) reduces the block size by $79\%$ $\left(\frac{25-120}{120}\right)$ while only sacrificing {\color{black}$0.038$ $\left(0.878-0.840\right)$} in power. Similarly, when $p=6$, the design with $k=3$ and $n_B=12$ (Table~\ref{tb:12}) reduces the block size by $90\%$ $\left(\frac{12-120}{120}\right)$ with only a {\color{black}$0.048$ $\left(0.894-0.846\right)$} reduction in power.
The type I error rate generally remains around $0.05$ when {\color{black}$p \leq 2$} but increases as $p$ rises. Compared to full block OofA designs, the constructed designs show only a modest increase in the type I error rate. For instance, when $p=6$, the design with $k=2$ and $n_B=25$ ({\color{black} Appendix Table~\ref{tb:25}}) only raises the type I error rate by {\color{black}$0.014$ $\left(0.100 - 0.086\right)$}. For designs with $k=3$ and $n_B=12$ (Table~\ref{tb:12}), the increase is {\color{black}$0.015$ $\left(0.103 - 0.088\right)$} at $p=6$.
{\color{black}The average differences ($DIF$) between the responses of the true optimal order and the estimated optimal order are less than 0.389 and 0.472 when $k = 2$ and $k = 3$, respectively. These differences are relatively small compared with the random error $\sigma_\epsilon = 1$. 
These results indicate that the proposed block OofA designs have comparable power, low type I error rates, and that the responses of the estimated optimal orders are relatively close to those of the true optimal orders. 

To evaluate the performance of the proposed block OofA designs with a larger number of factors, we conduct simulations with $m = 8$ to examine the overall power, type I error rates, and the average response differences for $k = 2$ block OofA designs with sizes $n_B = 70, 56$ and $42$, when the number of active effects $p \leq 8$. The results are presented in Appendix Table~\ref{tb:m=8}, which shows that the proposed designs retain high power, low type I error rates, and small response differences between the true and estimated optimal orders for larger value of $m$.}

\section{Concluding remarks}\label{se:con}  
In this study, we aimed to address the need for efficient and robust block designs in OofA experiments. We developed a criterion for assessing block OofA designs and proposed algorithms for their construction, utilizing the indicator function. Our results suggest that the proposed designs achieve meaningful reductions in block sizes while maintaining comparable power and type I error rates.

A notable aspect of this work is the introduction of an indicator function tailored to position-based models for block OofA designs. This function offers a standardized approach for evaluating aliasing and confounding structures, as well as the overall efficiency of the designs. By quantifying the severity of aliasing and confounding, it helps to better manage experimental conditions, potentially leading to more accurate and reliable outcomes.

Our algorithms for selecting optimal block OofA designs have proven to be effective. They leverage the proposed criterion to identify designs that achieve a balance between reduced block sizes and high statistical efficiency. This balance is crucial in practical applications where resources are limited and experimental conditions vary. 
The success of our methods in simulation studies suggests they may be applicable across a range of experimental scenarios.

The principles and techniques developed here could extend beyond block OofA designs to other experimental contexts where the order of component addition is important. For example, fields such as chemical process optimization, clinical trials, and industrial manufacturing might benefit from the methodologies presented. By offering a framework for evaluating and constructing block designs, our work provides a foundation for future advancements in these areas.

In conclusion, the advancements presented in this study contribute to the field of OofA experiments by offering practical solutions to challenges posed by heterogeneous experimental units. Future research may explore applying these methods to more complex settings and incorporating additional constraints to further enhance design efficiency. As the field evolves, the approaches and findings from this study could serve as useful resources for researchers and practitioners seeking to optimize experimental designs in various contexts.

{\color{black}
\section*{Data Availability Statement}
The data that support the findings of this study are available from the corresponding author upon request.
}

\section*{Acknowledgements}
This work was supported by the National Science and Technology Council, Taiwan, under Grant NSTC 113-2118-M-005-004-MY3.

\begin{appendices}
\section{Appendix Tables}

\setcounter{table}{0}
\renewcommand{\thetable}{A\arabic{table}}

\begin{table}[H]
\centering
\caption{{\color{black}Notation glossary for the OofA and proposed block OofA design framework.}}
\label{tb:notation}
\renewcommand{\arraystretch}{0.69}
{\color{black}
\begin{tabular}{ll}
\hline
Notation & Description / Representation \\
\hline
$m$ & Number of components in the experiment \\
$k$ & Number of blocks in the experiment \\
$n$; $n_B$ & Total number of runs; number of runs per block \\
$b$ & Block index, $b = 1, \ldots, k$ \\
$Z_j$ & Position factor of the $j$th component in the experiment \\
$B$ & Block factor in the block OofA design \\
$\bz=(z_1,\cdots,z_m)$ & Design point, a permutation of the positions $\{1, 2, \cdots, m\}$\\
$\bz'=(z_1,\cdots,z_m,b)$ & Design point in the block OofA design including block index $b$\\
$\cal O$; $\cal O'$ & Full OofA and block OofA designs \\
$\cal D$; $\cal D'$ & Specific OofA design and block OofA design selected from $\cal O$ and $\cal O'$ \\
$p_u(z); c_u(b)$ & Orthogonal polynomial contrasts of degree $u$ for position $z$ and block $b$\\
$T$; $S$  & Sets of polynomial degrees for position effects and block effects \\
${\cal T}$; ${\cal T'}$ & ${\cal T}=T^m$; ${\cal T'}=T^m\times S$ (for block OofA designs)\\

${\bt}$; $\bt'$ & $\bt= (t_1, \cdots, t_m)\in \cal T$;  $\bt'=(t_1,t_2,\cdots,t_m,s)\in{\cal T'}$\\ 

$X_\bt(\bz)$; $X_{\bt'}(\bz')$ & Polynomial terms in $F_{\cal D}(\bz)$ and $F_{\cal D'}(\bz')$\\
$F_{\cal D}(\bz)$; $F_{\cal D'}(\bz')$ & Indicator functions for an OofA design $\cal D$ and a block OofA design $\cal D'$\\
$a_\bt$; $a_{\bt'}$ & Coefficients of terms $X_\bt(\bz)$ and $X_{\bt'}(\bz')$ in the indicator functions \\
$||\bt||$; $||\bt'||_B$ & Polynomial degrees of $\bt$ and $(t_1, \ldots, t_m)$ in $\bt'$\\
$w_l$ & Aliasing of $l$th-order position effects in an OofA design\\ 
$w^P$; $w^B$ & Pure-type and mixed-type words for block OofA designs\\
$w_l^P$; $w_l^B$ & Aliasing and confounding of $l$th-order position effects in a block OofA design\\ 
$w_{(i)}'$ & $w_{(i)}'=w_l^P$ when $i=2(l-1)+1$ and $w_{(i)}'=w_l^B$ when $i=2(l-1)+2$\\
$W$; $W'$ & Word length patterns (WLPs) of OofA and block OofA designs\\
$L$; $L_i$ & $L$ is a set of Latin squares (LSs) $L_i$ for $i=1,\cdots,(m-1)!$\\ 
$C$; $C_i$ & $C$ is a set of component orthogonal arrays (COAs) $C_i$ for $i=1,\cdots,(m-2)!$\\
$\lambda$; $\gamma$; $\delta$ & $\lambda = \left\lfloor \frac{n_B}{m(m-1)} \right\rfloor$; $\gamma = \left\lfloor \frac{n_B - \lambda m(m-1)}{m} \right\rfloor$; $\delta = n_B - \lambda m(m-1) - \gamma m$\\


$\beta_0$; $\beta_j$; $\beta_{jj}$; $\beta_{jl}$ & Model coefficients for mean, linear, quadratic, and interaction effects \\
$\sigma_\epsilon$ & Standard deviation of the random error $\epsilon$\\
$PW$; $TY1$; $DIF$ & Power, type I error rate, and response difference metric \\

\hline
\end{tabular}
}
\end{table}


\begin{spacing}{1} 

\begin{table}[H]
\begin{center}
\caption{24 Latin squares for $m=5$}\label{tb:ls}
\begin{tabular}{cccc}									
\begin{tabular}{ccccc}									
\multicolumn{5}{c}{$L_1$}\\									
\hline									
1	&	2	&	3	&	4	&	5	\\
2	&	3	&	4	&	5	&	1	\\
3	&	4	&	5	&	1	&	2	\\
4	&	5	&	1	&	2	&	3	\\
5	&	1	&	2	&	3	&	4	\\
\hline									
\end{tabular}									
&									
\begin{tabular}{ccccc}									
\multicolumn{5}{c}{$L_2$}\\									
\hline									
1	&	3	&	5	&	2	&	4	\\
2	&	4	&	1	&	3	&	5	\\
3	&	5	&	2	&	4	&	1	\\
4	&	1	&	3	&	5	&	2	\\
5	&	2	&	4	&	1	&	3	\\
\hline									
\end{tabular}									
&									
\begin{tabular}{ccccc}									
\multicolumn{5}{c}{$L_3$}\\									
\hline									
1	&	4	&	2	&	5	&	3	\\
2	&	5	&	3	&	1	&	4	\\
3	&	1	&	4	&	2	&	5	\\
4	&	2	&	5	&	3	&	1	\\
5	&	3	&	1	&	4	&	2	\\
\hline									
\end{tabular}									
&									
\begin{tabular}{ccccc}									
\multicolumn{5}{c}{$L_4$}\\									
\hline									
1	&	5	&	4	&	3	&	2	\\
2	&	1	&	5	&	4	&	3	\\
3	&	2	&	1	&	5	&	4	\\
4	&	3	&	2	&	1	&	5	\\
5	&	4	&	3	&	2	&	1	\\
\hline									
\end{tabular}\\									
\multicolumn{4}{c}{}\\									
\begin{tabular}{ccccc}									
\multicolumn{5}{c}{$L_5$}\\									
\hline									
1	&	2	&	3	&	5	&	4	\\
2	&	3	&	4	&	1	&	5	\\
3	&	4	&	5	&	2	&	1	\\
4	&	5	&	1	&	3	&	2	\\
5	&	1	&	2	&	4	&	3	\\
\hline									
\end{tabular}									
&									
\begin{tabular}{ccccc}									
\multicolumn{5}{c}{$L_6$}\\									
\hline									
1	&	3	&	5	&	4	&	2	\\
2	&	4	&	1	&	5	&	3	\\
3	&	5	&	2	&	1	&	4	\\
4	&	1	&	3	&	2	&	5	\\
5	&	2	&	4	&	3	&	1	\\
\hline									
\end{tabular}									
&									
\begin{tabular}{ccccc}									
\multicolumn{5}{c}{$L_7$}\\									
\hline									
1	&	4	&	2	&	3	&	5	\\
2	&	5	&	3	&	4	&	1	\\
3	&	1	&	4	&	5	&	2	\\
4	&	2	&	5	&	1	&	3	\\
5	&	3	&	1	&	2	&	4	\\
\hline									
\end{tabular}									
&									
\begin{tabular}{ccccc}									
\multicolumn{5}{c}{$L_8$}\\									
\hline									
1	&	5	&	4	&	2	&	3	\\
2	&	1	&	5	&	3	&	4	\\
3	&	2	&	1	&	4	&	5	\\
4	&	3	&	2	&	5	&	1	\\
5	&	4	&	3	&	1	&	2	\\
\hline									
\end{tabular}\\									
\multicolumn{4}{c}{}\\									
\begin{tabular}{ccccc}									
\multicolumn{5}{c}{$L_9$}\\									
\hline									
1	&	2	&	4	&	3	&	5	\\
2	&	3	&	5	&	4	&	1	\\
3	&	4	&	1	&	5	&	2	\\
4	&	5	&	2	&	1	&	3	\\
5	&	1	&	3	&	2	&	4	\\
\hline									
\end{tabular}									
&									
\begin{tabular}{ccccc}									
\multicolumn{5}{c}{$L_{10}$}\\									
\hline									
1	&	3	&	2	&	5	&	4	\\
2	&	4	&	3	&	1	&	5	\\
3	&	5	&	4	&	2	&	1	\\
4	&	1	&	5	&	3	&	2	\\
5	&	2	&	1	&	4	&	3	\\
\hline									
\end{tabular}									
&									
\begin{tabular}{ccccc}									
\multicolumn{5}{c}{$L_{11}$}\\									
\hline									
1	&	4	&	5	&	2	&	3	\\
2	&	5	&	1	&	3	&	4	\\
3	&	1	&	2	&	4	&	5	\\
4	&	2	&	3	&	5	&	1	\\
5	&	3	&	4	&	1	&	2	\\
\hline									
\end{tabular}									
&									
\begin{tabular}{ccccc}									
\multicolumn{5}{c}{$L_{12}$}\\									
\hline									
1	&	5	&	3	&	4	&	2	\\
2	&	1	&	4	&	5	&	3	\\
3	&	2	&	5	&	1	&	4	\\
4	&	3	&	1	&	2	&	5	\\
5	&	4	&	2	&	3	&	1	\\
\hline									
\end{tabular}\\									
\multicolumn{4}{c}{}\\									
\begin{tabular}{ccccc}									
\multicolumn{5}{c}{$L_{13}$}\\									
\hline									
1	&	2	&	4	&	5	&	3	\\
2	&	3	&	5	&	1	&	4	\\
3	&	4	&	1	&	2	&	5	\\
4	&	5	&	2	&	3	&	1	\\
5	&	1	&	3	&	4	&	2	\\
\hline									
\end{tabular}									
&									
\begin{tabular}{ccccc}									
\multicolumn{5}{c}{$L_{14}$}\\									
\hline									
1	&	3	&	2	&	4	&	5	\\
2	&	4	&	3	&	5	&	1	\\
3	&	5	&	4	&	1	&	2	\\
4	&	1	&	5	&	2	&	3	\\
5	&	2	&	1	&	3	&	4	\\
\hline									
\end{tabular}									
&									
\begin{tabular}{ccccc}									
\multicolumn{5}{c}{$L_{15}$}\\									
\hline									
1	&	4	&	5	&	3	&	2	\\
2	&	5	&	1	&	4	&	3	\\
3	&	1	&	2	&	5	&	4	\\
4	&	2	&	3	&	1	&	5	\\
5	&	3	&	4	&	2	&	1	\\
\hline									
\end{tabular}									
&									
\begin{tabular}{ccccc}									
\multicolumn{5}{c}{$L_{16}$}\\									
\hline									
1	&	5	&	3	&	2	&	4	\\
2	&	1	&	4	&	3	&	5	\\
3	&	2	&	5	&	4	&	1	\\
4	&	3	&	1	&	5	&	2	\\
5	&	4	&	2	&	1	&	3	\\
\hline									
\end{tabular}\\									
\multicolumn{4}{c}{}\\									
\begin{tabular}{ccccc}									
\multicolumn{5}{c}{$L_{17}$}\\									
\hline									
1	&	2	&	5	&	3	&	4	\\
2	&	3	&	1	&	4	&	5	\\
3	&	4	&	2	&	5	&	1	\\
4	&	5	&	3	&	1	&	2	\\
5	&	1	&	4	&	2	&	3	\\
\hline									
\end{tabular}									
&									
\begin{tabular}{ccccc}									
\multicolumn{5}{c}{$L_{18}$}\\									
\hline									
1	&	3	&	4	&	5	&	2	\\
2	&	4	&	5	&	1	&	3	\\
3	&	5	&	1	&	2	&	4	\\
4	&	1	&	2	&	3	&	5	\\
5	&	2	&	3	&	4	&	1	\\
\hline									
\end{tabular}									
&									
\begin{tabular}{ccccc}									
\multicolumn{5}{c}{$L_{19}$}\\									
\hline									
1	&	4	&	3	&	2	&	5	\\
2	&	5	&	4	&	3	&	1	\\
3	&	1	&	5	&	4	&	2	\\
4	&	2	&	1	&	5	&	3	\\
5	&	3	&	2	&	1	&	4	\\
\hline									
\end{tabular}									
&									
\begin{tabular}{ccccc}									
\multicolumn{5}{c}{$L_{20}$}\\									
\hline									
1	&	5	&	2	&	4	&	3	\\
2	&	1	&	3	&	5	&	4	\\
3	&	2	&	4	&	1	&	5	\\
4	&	3	&	5	&	2	&	1	\\
5	&	4	&	1	&	3	&	2	\\
\hline									
\end{tabular}\\									
\multicolumn{4}{c}{}\\									
\begin{tabular}{ccccc}									
\multicolumn{5}{c}{$L_{21}$}\\									
\hline									
1	&	2	&	5	&	4	&	3	\\
2	&	3	&	1	&	5	&	4	\\
3	&	4	&	2	&	1	&	5	\\
4	&	5	&	3	&	2	&	1	\\
5	&	1	&	4	&	3	&	2	\\
\hline									
\end{tabular}									
&									
\begin{tabular}{ccccc}									
\multicolumn{5}{c}{$L_{22}$}\\									
\hline									
1	&	3	&	4	&	2	&	5	\\
2	&	4	&	5	&	3	&	1	\\
3	&	5	&	1	&	4	&	2	\\
4	&	1	&	2	&	5	&	3	\\
5	&	2	&	3	&	1	&	4	\\
\hline									
\end{tabular}									
&									
\begin{tabular}{ccccc}									
\multicolumn{5}{c}{$L_{23}$}\\									
\hline									
1	&	4	&	3	&	5	&	2	\\
2	&	5	&	4	&	1	&	3	\\
3	&	1	&	5	&	2	&	4	\\
4	&	2	&	1	&	3	&	5	\\
5	&	3	&	2	&	4	&	1	\\
\hline									
\end{tabular}									
&									
\begin{tabular}{ccccc}									
\multicolumn{5}{c}{$L_{24}$}\\									
\hline									
1	&	5	&	2	&	3	&	4	\\
2	&	1	&	3	&	4	&	5	\\
3	&	2	&	4	&	5	&	1	\\
4	&	3	&	5	&	1	&	2	\\
5	&	4	&	1	&	2	&	3	\\
\hline									
\end{tabular}\\									
\end{tabular}									
\end{center}
\end{table}

\begin{table}[H]
\begin{center}
\caption{Block OofA design for $m=5$, $k=2$, and $n_B=27$}\label{tb:27}
\begin{tabular}{ccc}													
\begin{tabular}{ccccccc}													
\hline													
Run	&	$Z_1$	&	$Z_2$	&	$Z_3$	&	$Z_4$	&	$Z_5$	&	$B$	\\
\hline													
1	&	1	&	2	&	3	&	4	&	5	&	1	\\
2	&	2	&	3	&	4	&	5	&	1	&	1	\\
3	&	3	&	4	&	5	&	1	&	2	&	1	\\
4	&	4	&	5	&	1	&	2	&	3	&	1	\\
5	&	5	&	1	&	2	&	3	&	4	&	1	\\
6	&	1	&	3	&	5	&	2	&	4	&	1	\\
7	&	2	&	4	&	1	&	3	&	5	&	1	\\
8	&	3	&	5	&	2	&	4	&	1	&	1	\\
9	&	4	&	1	&	3	&	5	&	2	&	1	\\
10	&	5	&	2	&	4	&	1	&	3	&	1	\\
11	&	1	&	4	&	2	&	5	&	3	&	1	\\
12	&	2	&	5	&	3	&	1	&	4	&	1	\\
13	&	3	&	1	&	4	&	2	&	5	&	1	\\
14	&	4	&	2	&	5	&	3	&	1	&	1	\\
15	&	5	&	3	&	1	&	4	&	2	&	1	\\
16	&	1	&	5	&	4	&	3	&	2	&	1	\\
17	&	2	&	1	&	5	&	4	&	3	&	1	\\
18	&	3	&	2	&	1	&	5	&	4	&	1	\\
19	&	4	&	3	&	2	&	1	&	5	&	1	\\
20	&	5	&	4	&	3	&	2	&	1	&	1	\\
21	&	1	&	2	&	4	&	3	&	5	&	1	\\
22	&	2	&	3	&	5	&	4	&	1	&	1	\\
23	&	3	&	4	&	1	&	5	&	2	&	1	\\
24	&	4	&	5	&	2	&	1	&	3	&	1	\\
25	&	5	&	1	&	3	&	2	&	4	&	1	\\
26	&	2	&	4	&	3	&	1	&	5	&	1	\\
27	&	4	&	1	&	5	&	3	&	2	&	1	\\
\hline													
\end{tabular}													
&													
\begin{tabular}{ccccccc}													
\hline													
Run	&	$Z_1$	&	$Z_2$	&	$Z_3$	&	$Z_4$	&	$Z_5$	&	$B$	\\
\hline													
28	&	1	&	2	&	3	&	5	&	4	&	2	\\
29	&	2	&	3	&	4	&	1	&	5	&	2	\\
30	&	3	&	4	&	5	&	2	&	1	&	2	\\
31	&	4	&	5	&	1	&	3	&	2	&	2	\\
32	&	5	&	1	&	2	&	4	&	3	&	2	\\
33	&	1	&	3	&	5	&	4	&	2	&	2	\\
34	&	2	&	4	&	1	&	5	&	3	&	2	\\
35	&	3	&	5	&	2	&	1	&	4	&	2	\\
36	&	4	&	1	&	3	&	2	&	5	&	2	\\
37	&	5	&	2	&	4	&	3	&	1	&	2	\\
38	&	1	&	4	&	2	&	3	&	5	&	2	\\
39	&	2	&	5	&	3	&	4	&	1	&	2	\\
40	&	3	&	1	&	4	&	5	&	2	&	2	\\
41	&	4	&	2	&	5	&	1	&	3	&	2	\\
42	&	5	&	3	&	1	&	2	&	4	&	2	\\
43	&	1	&	5	&	4	&	2	&	3	&	2	\\
44	&	2	&	1	&	5	&	3	&	4	&	2	\\
45	&	3	&	2	&	1	&	4	&	5	&	2	\\
46	&	4	&	3	&	2	&	5	&	1	&	2	\\
47	&	5	&	4	&	3	&	1	&	2	&	2	\\
48	&	1	&	4	&	5	&	2	&	3	&	2	\\
49	&	2	&	5	&	1	&	3	&	4	&	2	\\
50	&	3	&	1	&	2	&	4	&	5	&	2	\\
51	&	4	&	2	&	3	&	5	&	1	&	2	\\
52	&	5	&	3	&	4	&	1	&	2	&	2	\\
53	&	5	&	2	&	1	&	4	&	3	&	2	\\
54	&	3	&	5	&	4	&	2	&	1	&	2	\\
\hline													
\end{tabular}													
\end{tabular}													
\end{center}
\end{table}

\begin{table}[H]
\begin{center}
\caption{Block OofA design for $m=5$, $k=3$, and $n_B=15$}\label{tb:15}
\begin{tabular}{ccc}													
\begin{tabular}{cc@{ }c@{ }c@{ }c@{ }c@{ }c}													
\hline													
Run	&	$Z_1$	&	$Z_2$	&	$Z_3$	&	$Z_4$	&	$Z_5$	&	$B$	\\
\hline													
1	&	1	&	2	&	3	&	4	&	5	&	1	\\
2	&	2	&	3	&	4	&	5	&	1	&	1	\\
3	&	3	&	4	&	5	&	1	&	2	&	1	\\
4	&	4	&	5	&	1	&	2	&	3	&	1	\\
5	&	5	&	1	&	2	&	3	&	4	&	1	\\
6	&	1	&	5	&	4	&	3	&	2	&	1	\\
7	&	2	&	1	&	5	&	4	&	3	&	1	\\
8	&	3	&	2	&	1	&	5	&	4	&	1	\\
9	&	4	&	3	&	2	&	1	&	5	&	1	\\
10	&	5	&	4	&	3	&	2	&	1	&	1	\\
11	&	1	&	4	&	2	&	5	&	3	&	1	\\
12	&	2	&	5	&	3	&	1	&	4	&	1	\\
13	&	3	&	1	&	4	&	2	&	5	&	1	\\
14	&	4	&	2	&	5	&	3	&	1	&	1	\\
15	&	5	&	3	&	1	&	4	&	2	&	1	\\
\hline													
\end{tabular}													
&													
\begin{tabular}{cc@{ }c@{ }c@{ }c@{ }c@{ }c}													
\hline													
Run	&	$Z_1$	&	$Z_2$	&	$Z_3$	&	$Z_4$	&	$Z_5$	&	$B$	\\
\hline													
16	&	1	&	4	&	2	&	3	&	5	&	2	\\
17	&	2	&	5	&	3	&	4	&	1	&	2	\\
18	&	3	&	1	&	4	&	5	&	2	&	2	\\
19	&	4	&	2	&	5	&	1	&	3	&	2	\\
20	&	5	&	3	&	1	&	2	&	4	&	2	\\
21	&	1	&	5	&	4	&	2	&	3	&	2	\\
22	&	2	&	1	&	5	&	3	&	4	&	2	\\
23	&	3	&	2	&	1	&	4	&	5	&	2	\\
24	&	4	&	3	&	2	&	5	&	1	&	2	\\
25	&	5	&	4	&	3	&	1	&	2	&	2	\\
26	&	1	&	2	&	4	&	3	&	5	&	2	\\
27	&	2	&	3	&	5	&	4	&	1	&	2	\\
28	&	3	&	4	&	1	&	5	&	2	&	2	\\
29	&	4	&	5	&	2	&	1	&	3	&	2	\\
30	&	5	&	1	&	3	&	2	&	4	&	2	\\
\hline													
\end{tabular}													
&													
\begin{tabular}{cc@{ }c@{ }c@{ }c@{ }c@{ }c}													
\hline													
Run	&	$Z_1$	&	$Z_2$	&	$Z_3$	&	$Z_4$	&	$Z_5$	&	$B$	\\
\hline													
31	&	1	&	2	&	3	&	5	&	4	&	3	\\
32	&	2	&	3	&	4	&	1	&	5	&	3	\\
33	&	3	&	4	&	5	&	2	&	1	&	3	\\
34	&	4	&	5	&	1	&	3	&	2	&	3	\\
35	&	5	&	1	&	2	&	4	&	3	&	3	\\
36	&	1	&	3	&	5	&	2	&	4	&	3	\\
37	&	2	&	4	&	1	&	3	&	5	&	3	\\
38	&	3	&	5	&	2	&	4	&	1	&	3	\\
39	&	4	&	1	&	3	&	5	&	2	&	3	\\
40	&	5	&	2	&	4	&	1	&	3	&	3	\\
41	&	1	&	3	&	5	&	4	&	2	&	3	\\
42	&	2	&	4	&	1	&	5	&	3	&	3	\\
43	&	3	&	5	&	2	&	1	&	4	&	3	\\
44	&	4	&	1	&	3	&	2	&	5	&	3	\\
45	&	5	&	2	&	4	&	3	&	1	&	3	\\
\hline													
\end{tabular}													
\end{tabular}																										
\end{center}
\end{table}

\begin{table}[H]
\begin{center}
\caption{Block OofA design for $m=5$, $k=2$, and $n_B=25$}\label{tb:25}
\begin{tabular}{ccc}													
\begin{tabular}{ccccccc}													
\hline													
Run	&	$Z_1$	&	$Z_2$	&	$Z_3$	&	$Z_4$	&	$Z_5$	&	$B$	\\
\hline													
1	&	1	&	2	&	3	&	4	&	5	&	1	\\
2	&	2	&	3	&	4	&	5	&	1	&	1	\\
3	&	3	&	4	&	5	&	1	&	2	&	1	\\
4	&	4	&	5	&	1	&	2	&	3	&	1	\\
5	&	5	&	1	&	2	&	3	&	4	&	1	\\
6	&	1	&	3	&	5	&	2	&	4	&	1	\\
7	&	2	&	4	&	1	&	3	&	5	&	1	\\
8	&	3	&	5	&	2	&	4	&	1	&	1	\\
9	&	4	&	1	&	3	&	5	&	2	&	1	\\
10	&	5	&	2	&	4	&	1	&	3	&	1	\\
11	&	1	&	4	&	2	&	5	&	3	&	1	\\
12	&	2	&	5	&	3	&	1	&	4	&	1	\\
13	&	3	&	1	&	4	&	2	&	5	&	1	\\
14	&	4	&	2	&	5	&	3	&	1	&	1	\\
15	&	5	&	3	&	1	&	4	&	2	&	1	\\
16	&	1	&	5	&	4	&	3	&	2	&	1	\\
17	&	2	&	1	&	5	&	4	&	3	&	1	\\
18	&	3	&	2	&	1	&	5	&	4	&	1	\\
19	&	4	&	3	&	2	&	1	&	5	&	1	\\
20	&	5	&	4	&	3	&	2	&	1	&	1	\\
21	&	1	&	3	&	2	&	5	&	4	&	1	\\
22	&	2	&	4	&	3	&	1	&	5	&	1	\\
23	&	3	&	5	&	4	&	2	&	1	&	1	\\
24	&	4	&	1	&	5	&	3	&	2	&	1	\\
25	&	5	&	2	&	1	&	4	&	3	&	1	\\
\hline													
\end{tabular}													
&													
\begin{tabular}{ccccccc}													
\hline													
Run	&	$Z_1$	&	$Z_2$	&	$Z_3$	&	$Z_4$	&	$Z_5$	&	$B$	\\
\hline													
26	&	1	&	2	&	3	&	5	&	4	&	2	\\
27	&	2	&	3	&	4	&	1	&	5	&	2	\\
28	&	3	&	4	&	5	&	2	&	1	&	2	\\
29	&	4	&	5	&	1	&	3	&	2	&	2	\\
30	&	5	&	1	&	2	&	4	&	3	&	2	\\
31	&	1	&	3	&	5	&	4	&	2	&	2	\\
32	&	2	&	4	&	1	&	5	&	3	&	2	\\
33	&	3	&	5	&	2	&	1	&	4	&	2	\\
34	&	4	&	1	&	3	&	2	&	5	&	2	\\
35	&	5	&	2	&	4	&	3	&	1	&	2	\\
36	&	1	&	4	&	2	&	3	&	5	&	2	\\
37	&	2	&	5	&	3	&	4	&	1	&	2	\\
38	&	3	&	1	&	4	&	5	&	2	&	2	\\
39	&	4	&	2	&	5	&	1	&	3	&	2	\\
40	&	5	&	3	&	1	&	2	&	4	&	2	\\
41	&	1	&	5	&	4	&	2	&	3	&	2	\\
42	&	2	&	1	&	5	&	3	&	4	&	2	\\
43	&	3	&	2	&	1	&	4	&	5	&	2	\\
44	&	4	&	3	&	2	&	5	&	1	&	2	\\
45	&	5	&	4	&	3	&	1	&	2	&	2	\\
46	&	1	&	2	&	4	&	3	&	5	&	2	\\
47	&	2	&	3	&	5	&	4	&	1	&	2	\\
48	&	3	&	4	&	1	&	5	&	2	&	2	\\
49	&	4	&	5	&	2	&	1	&	3	&	2	\\
50	&	5	&	1	&	3	&	2	&	4	&	2	\\
\hline													
\end{tabular}													
\end{tabular}													
\end{center}
\end{table}

\begin{table}[H]
\begin{center}
\caption{Block OofA design for $m=5$, $k=3$, and $n_B=20$}\label{tb:20}
\begin{tabular}{ccc}													
\begin{tabular}{cc@{ }c@{ }c@{ }c@{ }c@{ }c}													
\hline													
Run	&	$Z_1$	&	$Z_2$	&	$Z_3$	&	$Z_4$	&	$Z_5$	&	$B$	\\
\hline													
1	&	1	&	2	&	3	&	4	&	5	&	1	\\
2	&	2	&	3	&	4	&	5	&	1	&	1	\\
3	&	3	&	4	&	5	&	1	&	2	&	1	\\
4	&	4	&	5	&	1	&	2	&	3	&	1	\\
5	&	5	&	1	&	2	&	3	&	4	&	1	\\
6	&	1	&	3	&	5	&	2	&	4	&	1	\\
7	&	2	&	4	&	1	&	3	&	5	&	1	\\
8	&	3	&	5	&	2	&	4	&	1	&	1	\\
9	&	4	&	1	&	3	&	5	&	2	&	1	\\
10	&	5	&	2	&	4	&	1	&	3	&	1	\\
11	&	1	&	4	&	2	&	5	&	3	&	1	\\
12	&	2	&	5	&	3	&	1	&	4	&	1	\\
13	&	3	&	1	&	4	&	2	&	5	&	1	\\
14	&	4	&	2	&	5	&	3	&	1	&	1	\\
15	&	5	&	3	&	1	&	4	&	2	&	1	\\
16	&	1	&	5	&	4	&	3	&	2	&	1	\\
17	&	2	&	1	&	5	&	4	&	3	&	1	\\
18	&	3	&	2	&	1	&	5	&	4	&	1	\\
19	&	4	&	3	&	2	&	1	&	5	&	1	\\
20	&	5	&	4	&	3	&	2	&	1	&	1	\\
\hline													
\end{tabular}													
&													
\begin{tabular}{cc@{ }c@{ }c@{ }c@{ }c@{ }c}													
\hline													
Run	&	$Z_1$	&	$Z_2$	&	$Z_3$	&	$Z_4$	&	$Z_5$	&	$B$	\\
\hline													
21	&	1	&	2	&	3	&	5	&	4	&	2	\\
22	&	2	&	3	&	4	&	1	&	5	&	2	\\
23	&	3	&	4	&	5	&	2	&	1	&	2	\\
24	&	4	&	5	&	1	&	3	&	2	&	2	\\
25	&	5	&	1	&	2	&	4	&	3	&	2	\\
26	&	1	&	3	&	5	&	4	&	2	&	2	\\
27	&	2	&	4	&	1	&	5	&	3	&	2	\\
28	&	3	&	5	&	2	&	1	&	4	&	2	\\
29	&	4	&	1	&	3	&	2	&	5	&	2	\\
30	&	5	&	2	&	4	&	3	&	1	&	2	\\
31	&	1	&	4	&	2	&	3	&	5	&	2	\\
32	&	2	&	5	&	3	&	4	&	1	&	2	\\
33	&	3	&	1	&	4	&	5	&	2	&	2	\\
34	&	4	&	2	&	5	&	1	&	3	&	2	\\
35	&	5	&	3	&	1	&	2	&	4	&	2	\\
36	&	1	&	5	&	4	&	2	&	3	&	2	\\
37	&	2	&	1	&	5	&	3	&	4	&	2	\\
38	&	3	&	2	&	1	&	4	&	5	&	2	\\
39	&	4	&	3	&	2	&	5	&	1	&	2	\\
40	&	5	&	4	&	3	&	1	&	2	&	2	\\
\hline													
\end{tabular}													
&													
\begin{tabular}{cc@{ }c@{ }c@{ }c@{ }c@{ }c}													
\hline													
Run	&	$Z_1$	&	$Z_2$	&	$Z_3$	&	$Z_4$	&	$Z_5$	&	$B$	\\
\hline													
41	&	1	&	2	&	4	&	3	&	5	&	3	\\
42	&	2	&	3	&	5	&	4	&	1	&	3	\\
43	&	3	&	4	&	1	&	5	&	2	&	3	\\
44	&	4	&	5	&	2	&	1	&	3	&	3	\\
45	&	5	&	1	&	3	&	2	&	4	&	3	\\
46	&	1	&	3	&	2	&	5	&	4	&	3	\\
47	&	2	&	4	&	3	&	1	&	5	&	3	\\
48	&	3	&	5	&	4	&	2	&	1	&	3	\\
49	&	4	&	1	&	5	&	3	&	2	&	3	\\
50	&	5	&	2	&	1	&	4	&	3	&	3	\\
51	&	1	&	4	&	5	&	2	&	3	&	3	\\
52	&	2	&	5	&	1	&	3	&	4	&	3	\\
53	&	3	&	1	&	2	&	4	&	5	&	3	\\
54	&	4	&	2	&	3	&	5	&	1	&	3	\\
55	&	5	&	3	&	4	&	1	&	2	&	3	\\
56	&	1	&	5	&	3	&	4	&	2	&	3	\\
57	&	2	&	1	&	4	&	5	&	3	&	3	\\
58	&	3	&	2	&	5	&	1	&	4	&	3	\\
59	&	4	&	3	&	1	&	2	&	5	&	3	\\
60	&	5	&	4	&	2	&	3	&	1	&	3	\\
\hline													
\end{tabular}													
\end{tabular}													
\end{center}
\end{table}

\begin{table}[H]
\begin{center}
\caption{Block OofA design for $m=5$, $k=2$, and $n_B=40$}\label{tb:40}
\begin{tabular}{ccc}													
\begin{tabular}{ccccccc}													
\hline													
Run	&	$Z_1$	&	$Z_2$	&	$Z_3$	&	$Z_4$	&	$Z_5$	&	$B$	\\
\hline													
1	&	1	&	2	&	3	&	4	&	5	&	1	\\
2	&	2	&	3	&	4	&	5	&	1	&	1	\\
3	&	3	&	4	&	5	&	1	&	2	&	1	\\
4	&	4	&	5	&	1	&	2	&	3	&	1	\\
5	&	5	&	1	&	2	&	3	&	4	&	1	\\
6	&	1	&	3	&	5	&	2	&	4	&	1	\\
7	&	2	&	4	&	1	&	3	&	5	&	1	\\
8	&	3	&	5	&	2	&	4	&	1	&	1	\\
9	&	4	&	1	&	3	&	5	&	2	&	1	\\
10	&	5	&	2	&	4	&	1	&	3	&	1	\\
11	&	1	&	4	&	2	&	5	&	3	&	1	\\
12	&	2	&	5	&	3	&	1	&	4	&	1	\\
13	&	3	&	1	&	4	&	2	&	5	&	1	\\
14	&	4	&	2	&	5	&	3	&	1	&	1	\\
15	&	5	&	3	&	1	&	4	&	2	&	1	\\
16	&	1	&	5	&	4	&	3	&	2	&	1	\\
17	&	2	&	1	&	5	&	4	&	3	&	1	\\
18	&	3	&	2	&	1	&	5	&	4	&	1	\\
19	&	4	&	3	&	2	&	1	&	5	&	1	\\
20	&	5	&	4	&	3	&	2	&	1	&	1	\\
21	&	1	&	2	&	3	&	5	&	4	&	1	\\
22	&	2	&	3	&	4	&	1	&	5	&	1	\\
23	&	3	&	4	&	5	&	2	&	1	&	1	\\
24	&	4	&	5	&	1	&	3	&	2	&	1	\\
25	&	5	&	1	&	2	&	4	&	3	&	1	\\
26	&	1	&	3	&	5	&	4	&	2	&	1	\\
27	&	2	&	4	&	1	&	5	&	3	&	1	\\
28	&	3	&	5	&	2	&	1	&	4	&	1	\\
29	&	4	&	1	&	3	&	2	&	5	&	1	\\
30	&	5	&	2	&	4	&	3	&	1	&	1	\\
31	&	1	&	4	&	2	&	3	&	5	&	1	\\
32	&	2	&	5	&	3	&	4	&	1	&	1	\\
33	&	3	&	1	&	4	&	5	&	2	&	1	\\
34	&	4	&	2	&	5	&	1	&	3	&	1	\\
35	&	5	&	3	&	1	&	2	&	4	&	1	\\
36	&	1	&	5	&	4	&	2	&	3	&	1	\\
37	&	2	&	1	&	5	&	3	&	4	&	1	\\
38	&	3	&	2	&	1	&	4	&	5	&	1	\\
39	&	4	&	3	&	2	&	5	&	1	&	1	\\
40	&	5	&	4	&	3	&	1	&	2	&	1	\\
\hline													
\end{tabular}													
&													
\begin{tabular}{ccccccc}													
\hline													
Run	&	$Z_1$	&	$Z_2$	&	$Z_3$	&	$Z_4$	&	$Z_5$	&	$B$	\\
\hline													
41	&	1	&	2	&	4	&	3	&	5	&	2	\\
42	&	2	&	3	&	5	&	4	&	1	&	2	\\
43	&	3	&	4	&	1	&	5	&	2	&	2	\\
44	&	4	&	5	&	2	&	1	&	3	&	2	\\
45	&	5	&	1	&	3	&	2	&	4	&	2	\\
46	&	1	&	3	&	2	&	5	&	4	&	2	\\
47	&	2	&	4	&	3	&	1	&	5	&	2	\\
48	&	3	&	5	&	4	&	2	&	1	&	2	\\
49	&	4	&	1	&	5	&	3	&	2	&	2	\\
50	&	5	&	2	&	1	&	4	&	3	&	2	\\
51	&	1	&	4	&	5	&	2	&	3	&	2	\\
52	&	2	&	5	&	1	&	3	&	4	&	2	\\
53	&	3	&	1	&	2	&	4	&	5	&	2	\\
54	&	4	&	2	&	3	&	5	&	1	&	2	\\
55	&	5	&	3	&	4	&	1	&	2	&	2	\\
56	&	1	&	5	&	3	&	4	&	2	&	2	\\
57	&	2	&	1	&	4	&	5	&	3	&	2	\\
58	&	3	&	2	&	5	&	1	&	4	&	2	\\
59	&	4	&	3	&	1	&	2	&	5	&	2	\\
60	&	5	&	4	&	2	&	3	&	1	&	2	\\
61	&	1	&	2	&	4	&	5	&	3	&	2	\\
62	&	2	&	3	&	5	&	1	&	4	&	2	\\
63	&	3	&	4	&	1	&	2	&	5	&	2	\\
64	&	4	&	5	&	2	&	3	&	1	&	2	\\
65	&	5	&	1	&	3	&	4	&	2	&	2	\\
66	&	1	&	3	&	2	&	4	&	5	&	2	\\
67	&	2	&	4	&	3	&	5	&	1	&	2	\\
68	&	3	&	5	&	4	&	1	&	2	&	2	\\
69	&	4	&	1	&	5	&	2	&	3	&	2	\\
70	&	5	&	2	&	1	&	3	&	4	&	2	\\
71	&	1	&	4	&	5	&	3	&	2	&	2	\\
72	&	2	&	5	&	1	&	4	&	3	&	2	\\
73	&	3	&	1	&	2	&	5	&	4	&	2	\\
74	&	4	&	2	&	3	&	1	&	5	&	2	\\
75	&	5	&	3	&	4	&	2	&	1	&	2	\\
76	&	1	&	5	&	3	&	2	&	4	&	2	\\
77	&	2	&	1	&	4	&	3	&	5	&	2	\\
78	&	3	&	2	&	5	&	4	&	1	&	2	\\
79	&	4	&	3	&	1	&	5	&	2	&	2	\\
80	&	5	&	4	&	2	&	1	&	3	&	2	\\
\hline													
\end{tabular}													
\end{tabular}													
\end{center}
\end{table}

\begin{table}[H]
\begin{center}
\caption{\color{black}Practical cookbook for the required processes (COA: COA assignment; LSex: LS exchange; ROWex: row exchange) and the suggested default settings for the iteration parameters $(I_1, I_2, I_3)$, where ``-” indicates not applicable and ``A $+$ B” indicates that processes A and B are both required.}\label{tb:cook}
{\color{black}
\begin{tabular}{rrrcc}									
\hline									
$\lambda$	&	$\gamma$	&	$\delta$	&	Process	&	Suggested setting	\\
\hline									
0	&	0	&	0	&	-	&	-	\\
0	&	0	&	$>0$	&	ROWex	&	$(I_1,I_2,I_3)=(\lfloor500/m\rfloor,\text{-},k^2\delta^2)$	\\
0	&	$>0$	&	0	&	LSex 	&	$(I_1,I_2,I_3)=(\lfloor500/m\rfloor,k^2\gamma^2,\text{-})$	\\
0	&	$>0$	&	$>0$	&	LSex + ROWex	&	$(I_1,I_2,I_3)=(\lfloor500/m\rfloor,k^2\gamma^2,k^2\delta^2)$	\\
$>0$	&	0	&	0	&	COA  	&	-	\\
$>0$	&	0	&	$>0$	&	COA + ROWex	&	$(I_1,I_2,I_3)=(\lfloor500/m\rfloor,\text{-},k^2\delta^2)$	\\
$>0$	&	$>0$	&	0	&	COA + LSex 	&	$(I_1,I_2,I_3)=(\lfloor500/m\rfloor,k^2\gamma^2,\text{-})$	\\
$>0$	&	$>0$	&	$>0$	&	COA + LSex + ROWex	&	$(I_1,I_2,I_3)=(\lfloor500/m\rfloor,k^2\gamma^2,k^2\delta^2)$	\\
\hline									
\end{tabular}									}
\end{center}
\end{table}

\end{spacing}

\begin{spacing}{1}

\begin{landscape}

\begin{table}[H]
\begin{center}
\caption{\color{black}Forward regression with AIC for the experiment with the unblock OofA design in Table~\ref{tb:unbk}}\label{tb:aic_unbk}
{\color{black}
\begin{tabular}{crl}																					
\hline																					
Step	&	AIC	&	\multicolumn{1}{c}{Model}																	\\
\hline																					
1	&	144.0	&	$\hat y =	22.5															$	\\
2	&	121.6	&	$\hat y =	22.5	-5.1	Z_2^l													$	\\
3	&	113.4	&	$\hat y =	22.4	-5.1	Z_2^l	-2.5	Z_2^q											$	\\
4	&	109.4	&	$\hat y =	22.4	-4.3	Z_2^l	-2.5	Z_2^q	+	1.9	Z_5^l								$	\\
5	&	108.7	&	$\hat y =	22.1	-4.3	Z_2^l	-2.9	Z_2^q	+	1.9	Z_5^l	-1.3	Z_2^lZ_3^l						$	\\
6	&	108.1	&	$\hat y =	21.6	-4.3	Z_2^l	-3.4	Z_2^q	+	1.9	Z_5^l	-2.2	Z_2^lZ_3^l	-1.5	Z_1^lZ_2^l				$	\\
7	&	107.7	&	$\hat y =	21.6	-4.0	Z_2^l	-3.4	Z_2^q	+	2.3	Z_5^l	-2.2	Z_2^lZ_3^l	-1.5	Z_1^lZ_2^l	+	1.0	Z_1^l	$	\\
\hline																					
\end{tabular}																					
}
\end{center}
\end{table}

\begin{table}[H]
\begin{center}
\caption{\color{black}Forward regression with AIC for the experiment with the block OofA design in Table~\ref{tb:12}}\label{tb:aic_bk}
{\color{black}
\begin{tabular}{crl}																																	
\hline																																	
Step	&	AIC	&	\multicolumn{1}{c}{Model}																													\\
\hline																																	
1	&	144.0	&	$\hat y =	22.5																											$	\\
2	&	126.7	&	$\hat y =	22.5	-4.6	B^l																									$	\\
3	&	108.4	&	$\hat y =	22.5	-4.6	B^l	-3.7	Z_2^l																							$	\\
4	&	59.5	&	$\hat y =	22.4	-4.5	B^l	-3.7	Z_2^l	-3.6	Z_2^q																					$	\\
5	&	50.2	&	$\hat y =	22.4	-4.5	B^l	-3.7	Z_2^l	-3.4	Z_2^q	+	1.1	B^q																		$	\\
6	&	39.7	&	$\hat y =	22.4	-4.5	B^l	-3.2	Z_2^l	-3.4	Z_2^q	+	1.1	B^q	+	1.0	Z_5^l															$	\\
7	&	29.0	&	$\hat y =	22.8	-4.4	B^l	-3.2	Z_2^l	-3.1	Z_2^q	+	1.0	B^q	+	1.0	Z_5^l	+	1.1	Z_2^lZ_5^l												$	\\
8	&	15.8	&	$\hat y =	23.1	-4.3	B^l	-3.2	Z_2^l	-3.3	Z_2^q	+	1.1	B^q	+	1.0	Z_5^l	+	1.2	Z_2^lZ_5^l	+	0.9	Z_1^lZ_5^l									$	\\
9	&	5.6	&	$\hat y =	23.0	-4.4	B^l	-3.2	Z_2^l	-3.1	Z_2^q	+	1.0	B^q	+	1.0	Z_5^l	+	1.5	Z_2^lZ_5^l	+	1.0	Z_1^lZ_5^l	-0.7	Z_3^lZ_4^l							$	\\
10	&	3.8	&	$\hat y =	23.0	-4.4	B^l	-3.1	Z_2^l	-3.1	Z_2^q	+	1.0	B^q	+	1.1	Z_5^l	+	1.5	Z_2^lZ_5^l	+	1.0	Z_1^lZ_5^l	-0.7	Z_3^lZ_4^l	+	0.3	Z_1^l				$	\\
11	&	3.5	&	$\hat y =	23.0	-4.4	B^l	-3.1	Z_2^l	-3.0	Z_2^q	+	1.0	B^q	+	1.1	Z_5^l	+	1.4	Z_2^lZ_5^l	+	0.9	Z_1^lZ_5^l	-0.6	Z_3^lZ_4^l	+	0.3	Z_1^l	+	0.3	Z_4^q	$	\\
\hline																																	
\end{tabular}																																				
}
\end{center}
\end{table}
\end{landscape}

\begin{table}[H]
\begin{center}
\caption{{\color{black}Power ($PW$), type I error rate ($TY1$), and response difference between the true and estimated optimal orders ($DIF$) for $m=8$ and $k=2$ block OofA designs in the simulation study}}\label{tb:m=8}
{\color{black}
\begin{tabular}{cccccccccc}																			
\hline																			
$n_B$	&		&	$p=1$	&	$p=2$	&	$p=3$	&	$p=4$	&	$p=5$	&	$p=6$	&	$p=7$	&	$p=8$	\\
\hline																			
70	&	$PW$	&	1.000	&	1.000	&	1.000	&	0.999	&	0.972	&	0.966	&	0.975	&	0.975	\\
	&	$TY1$	&	0.045	&	0.046	&	0.044	&	0.042	&	0.047	&	0.052	&	0.051	&	0.054	\\
	&	$DIF$	&	0.531	&	0.518	&	0.408	&	0.382	&	0.361	&	0.340	&	0.368	&	0.396	\\
56	&	$PW$	&	1.000	&	1.000	&	1.000	&	1.000	&	0.980	&	0.978	&	0.981	&	0.984	\\
	&	$TY1$	&	0.046	&	0.045	&	0.043	&	0.044	&	0.047	&	0.048	&	0.049	&	0.051	\\
	&	$DIF$	&	0.630	&	0.562	&	0.436	&	0.407	&	0.421	&	0.386	&	0.403	&	0.485	\\
42	&	$PW$	&	1.000	&	1.000	&	1.000	&	0.998	&	0.969	&	0.956	&	0.975	&	0.967	\\
	&	$TY1$	&	0.044	&	0.042	&	0.042	&	0.044	&	0.050	&	0.052	&	0.052	&	0.059	\\
	&	$DIF$	&	0.696	&	0.642	&	0.526	&	0.490	&	0.501	&	0.479	&	0.521	&	0.688	\\
\hline																			
\end{tabular}	
}																		
\end{center}
\end{table}

\end{spacing}

\newpage
\section{Appendix Figure}

\setcounter{figure}{0}
\renewcommand{\thefigure}{A\arabic{figure}}

\begin{figure}[h]
\begin{subfigure}{.5\textwidth}
  \centering
  \includegraphics[width=.8\linewidth]{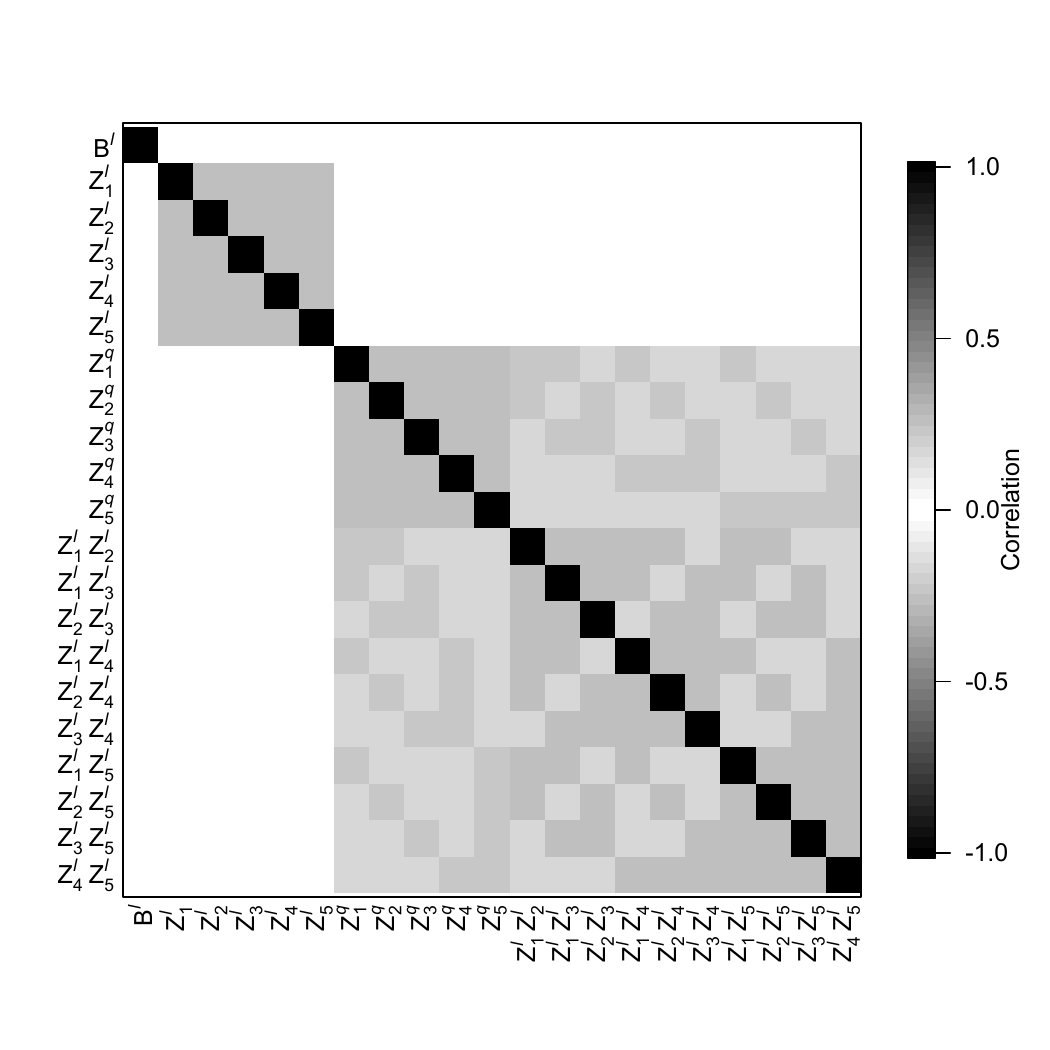}
  \caption{$n_B=120$ (Full design)}
  \label{fi:2x120}
\end{subfigure}%
\begin{subfigure}{.5\textwidth}
  \centering
  \includegraphics[width=.8\linewidth]{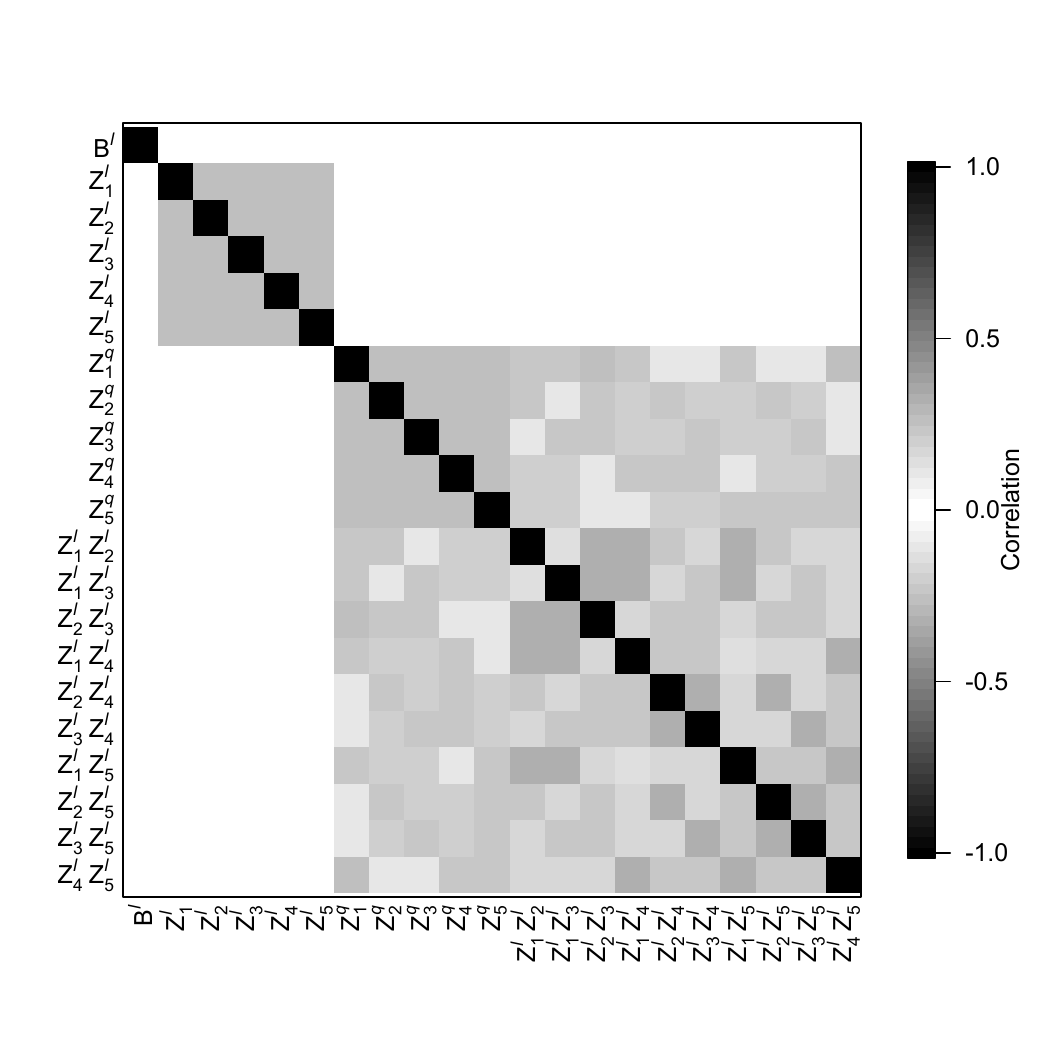}
  \caption{$n_B=40$ (Table \ref{tb:40})}  
  \label{fi:40}
\end{subfigure}
\begin{subfigure}{.5\textwidth}
  \centering
  \includegraphics[width=.8\linewidth]{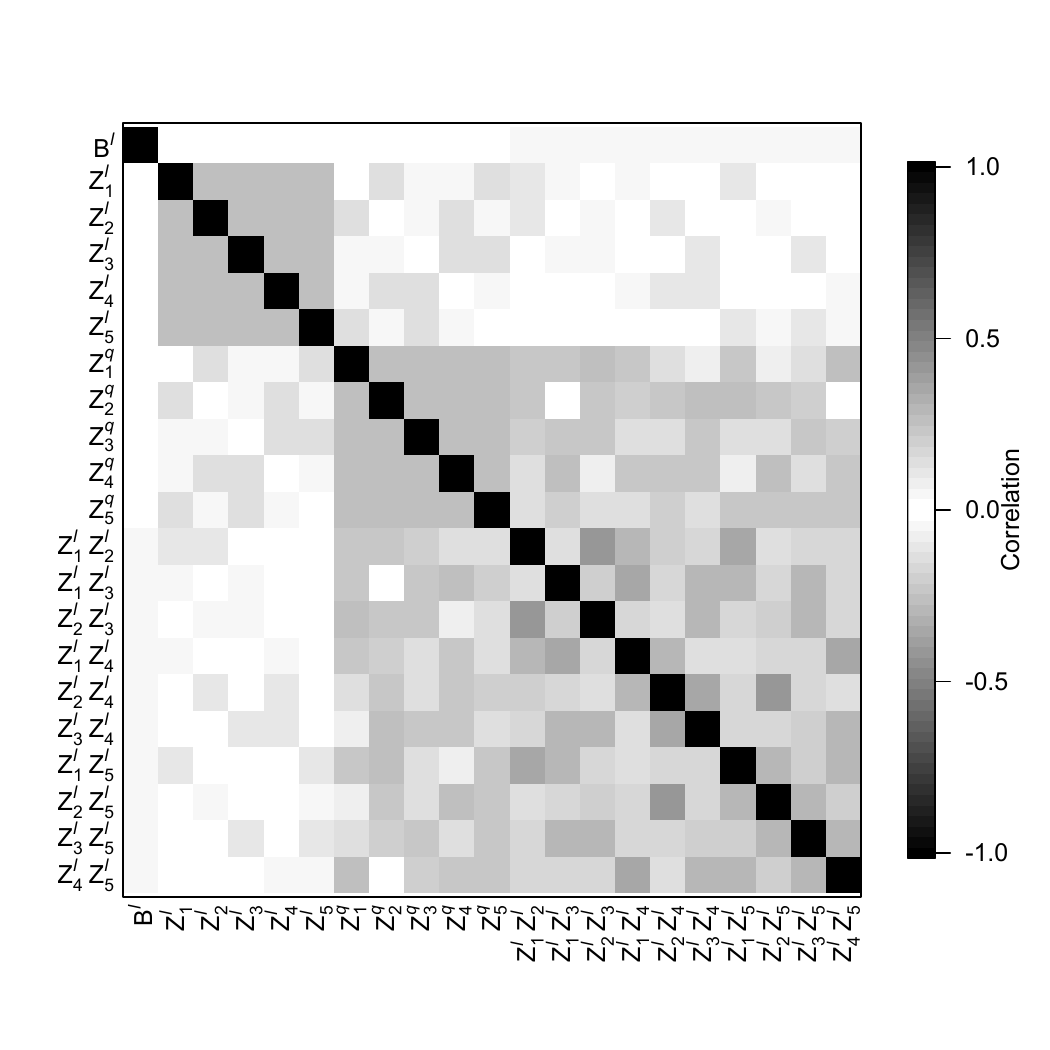}
  \caption{$n_B=25$ (Table \ref{tb:25})}
  \label{fi:25}
\end{subfigure}
\begin{subfigure}{.5\textwidth}
  \centering
  \includegraphics[width=.8\linewidth]{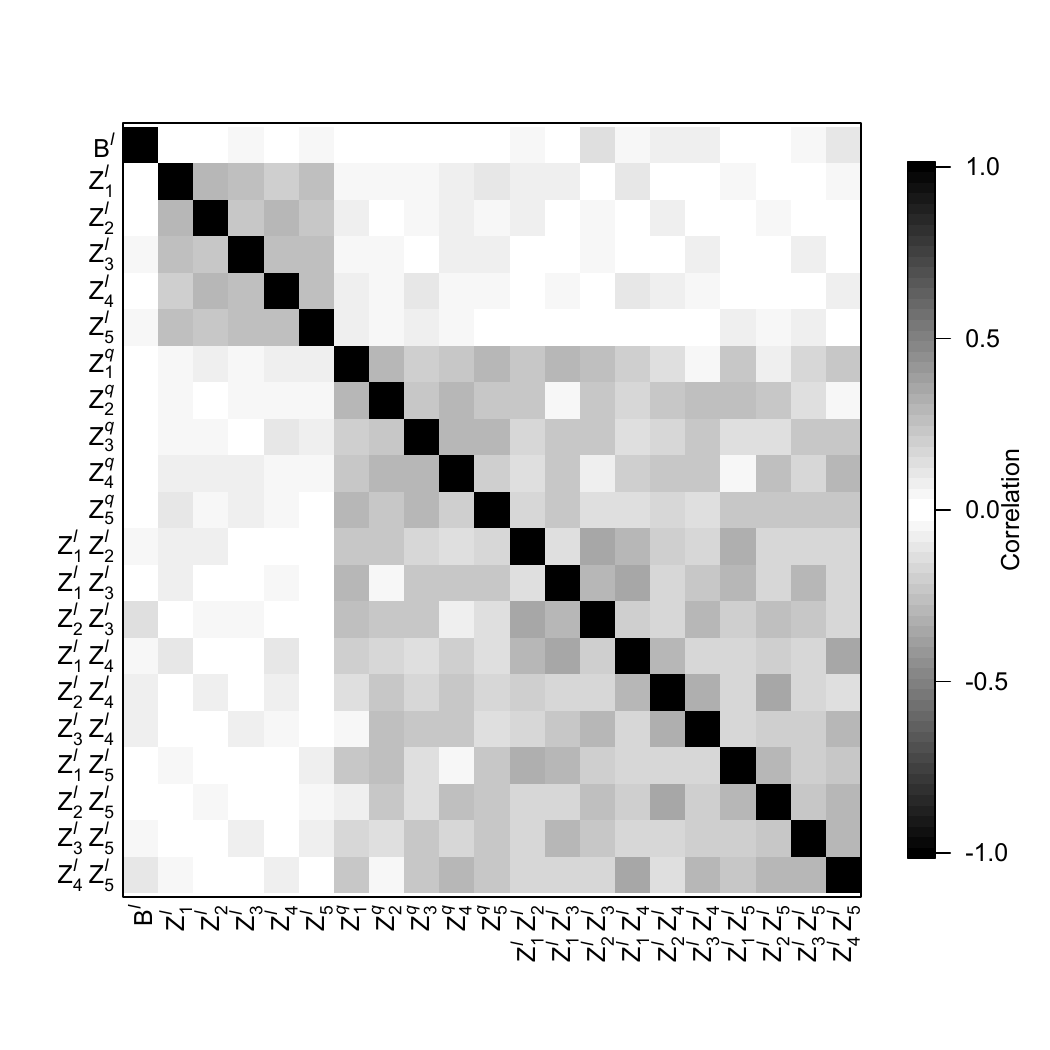}
  \caption{$n_B=27$  (Table \ref{tb:27})}
  \label{fi:27}
\end{subfigure}
\caption{Correlation plots of block OofA designs for $m=5,k=2$}
\label{fi:k2}
\end{figure}

\end{appendices}

\end{document}